\documentclass[a4paper,fleqn,usenatbib]{mnras}

\usepackage{newtxtext,newtxmath}

\usepackage[T1]{fontenc}

\DeclareRobustCommand{\VAN}[3]{#2}
\let\VANthebibliography\thebibliography
\def\thebibliography{\DeclareRobustCommand{\VAN}[3]{##3}\VANthebibliography}

\usepackage{graphicx} 
\usepackage{amsmath}	

\usepackage{dcolumn}
\usepackage{color}
\usepackage{subfigure}
\usepackage{float}
\usepackage{makecell}
\usepackage{enumitem}
\usepackage{hyperref}

\usepackage{mathrsfs}
\usepackage{bm}

\usepackage{enumitem}
\usepackage[english]{babel}
\usepackage{gensymb}
\usepackage{multirow}
\usepackage{booktabs}
\usepackage{threeparttable}
\usepackage{ulem}
\usepackage{hyperref}
\usepackage{soul}

\newcommand{\mr}{\mathrm}
\newcommand{\qvec}{\mathbfit}

\newcommand{\MHz}{\ensuremath{\, {\rm MHz}}}
\newcommand{\kHz}{\ensuremath{\, {\rm kHz}}}
\newcommand{\m}{\ensuremath{\, {\rm m}}}
\newcommand{\km}{\ensuremath{\, {\rm km}}}
\newcommand{\kms}{\ensuremath{\, {\rm km} s^{-1}}}
\newcommand{\K}{\ensuremath{\, {\rm K}}}

\title[DSL imaging]{Imaging sensitivity of a linear interferometer array on lunar orbit}

\author[Yuan Shi et al.]{Yuan Shi$^{1,2}$, Yidong Xu$^1$\thanks{E-mail: xuyd@nao.cas.cn (YX)},
Li Deng$^3$, Fengquan Wu$^1$, Lin Wu$^3$, Qizhi Huang$^1$, 
\newauthor
Shifan Zuo$^{4}$, Jingye Yan$^3$, Xuelei Chen$^{1,2,5,6}$\thanks{E-mail: xuelei@cosmology.bao.ac.cn (XC)}
\\
$^{1}$ National Astronomical Observatories, Chinese Academy of Sciences, 20A Datun Road, Beijing 100101, P. R. China\\
$^{2}$ University of Chinese Academy of Sciences, Beijing 100049, P. R. China\\
$^{3}$ National Space Science Center, Chinese Academy of Sciences, Beijing 100190, P. R. China\\
$^4$ Department of Astronomy and Tsinghua Center for Astrophysics, Tsinghua University, Beijing 100084, P. R. China\\
$^5$ Department of Physics, College of Sciences, Northeastern University, Shenyang 110819, China\\
$^6$ Center of High Energy Physics, Peking University, Beijing 100871, P. R. China}

\date{Accepted XXX. Received YYY; in original form ZZZ}

\pubyear{2021}

\begin{document}
\label{firstpage}
\pagerange{\pageref{firstpage}--\pageref{lastpage}}

\maketitle

\begin{abstract}
Ground-based observation at frequencies below 30 MHz is hindered by the ionosphere of the Earth and radio frequency interference. To map the sky at these low frequencies, we have proposed the Discovering the Sky at the Longest wavelength mission (DSL, also known as the ``Hongmeng'' mission, which means ``Primordial Universe'' in Chinese) concept, which employs a linear array of micro-satellites orbiting the Moon. Such an array can make interferometric observations achieving good angular resolutions despite the small size of the antennas. However, it differs from the conventional ground-based interferometer array or even the previous orbital interferometers in many aspects, new data-processing methods need to be developed. In this work, we make a series of simulations to assess the imaging quality and sensitivity of such an array. We start with an input sky model and a simple orbit model, generate mock interferometric visibilities, and then reconstruct the sky map. We consider various observational effects and practical issues, such as the system noise, antenna response, and Moon blockage. Based on the quality of the recovered image, we quantify the imaging capability of the array for different satellite numbers and array configurations. For the first time, we make practical estimates of the point source sensitivity for such a lunar orbit array, and predict the expected number of detectable sources for the mission. Depending on the radio source number distribution which is still very uncertain at these frequencies, the proposed mission can detect $10^2 \sim 10^4 $ sources during its operation. 

\end{abstract}



\section{Introduction}

Astronomy has entered the  ``multi-messenger'' era, with observations covering a broad range of wavelength for electromagnetic radiations, from the radio to the $\gamma$-ray, as well as non-electromagnetic signals such as gravitational wave, neutrino, and cosmic rays. In particular, low-frequency radio astronomy is undergoing a renaissance, with exciting prospects of measuring the redshifted 21 cm signals of neutral
hydrogen from the cosmic dawn and 
dark ages. However, the radio sky at frequencies below $\sim 30 \MHz$ or wavelength longer than $\sim 10$ m, is still largely unknown due the 
absorption and distortion by the Earth's ionosphere, and man-made and natural radio frequency interferences
(RFIs)\footnote{In the historical radio engineering parlance they belong to the ``high frequency"(HF) or "medium frequency" (MF) band, though in astronomy they are clearly at the lowest frequency band. Below we shall refer these as {\it ultra-long wavelength} to avoid confusion. Note that the interstellar medium is opaque to radio waves with frequency below $\sim 0.03 \MHz$, so the ULF band (0.3-3 kHz) is irrelevant to astronomical observation.}. 
Currently, there is scant data at such low frequencies, and the available data are old, often with a partial sky coverage, e.g. the Dominion Radio Astrophysical Observatory (DRAO) 22 MHz sky map \citep{roger1999radio}, 
or with very low resolutions, e.g. the data from the IMP-6
\citep{Brown1973} and RAE-2 \citep{Novaco1978,Cane1979} space missions.

In recent years, a number of	
space mission concepts have been proposed to make low frequency observations, and some are actively studied at present, such as the Discovering the Sky at the Longest wavelength (DSL; \citealt{2010cosp...38.2364B,2019arXiv190710853C,2020arXiv200715794C}), the
Dark Ages Polarimetry PathfindER (DAPPER; \citealt{2018JCAP...12..015T}), and the Farside Array for Radio Science Investigations of the Dark ages and Exoplanets (FARSIDE, \citealt{2019BAAS...51g.178B}).
By orbiting the Moon, or landing at the far-side of the Moon, the Moon is used as a natural shield against the RFI from the Earth, providing an ideal environment for such observation.

In the DSL or Hongmeng mission concept, an array of satellites  will be launched together as an assembly into the lunar orbit by a single rocket, then they will be released sequentially into a linear formation on the same circular orbit. One of the satellites will be a bigger one and shall be called the ``mother" satellite, which serves to collect the data from the other smaller ``daughter" satellites,
pre-process the data, and transmit the data back to the Earth with a high-gain antenna, while the daughter satellites make the actual observations. During the mission, the satellites will make both interferometric imaging and high precision global spectrum measurement on the part of orbit behind the Moon, and the scientific data will be transmitted to the Earth at the near side part of the orbit.

Although radio interferometry has long been used in astronomy, the lunar orbit array is very different from any previous or existing interferometer array, including even the orbital interferometers such as the HALCA \citep{2000PASJ...52..955H} or RadioAstron-A \citep{2013ARep...57..153K,2018arXiv181001230G}.  In the conceptual design of the DSL mission, the imaging daughter satellites are  each iequipped with a pair of deploy-able, orthogonally oriented tripole antennas to receive the radio signal, and when deployed, they will form effectively three pairs of split dipoles in three orthogonal directions (c.f. \citealt{2020arXiv200715794C}). As the observation is carried out at long wavelengths, for most of the observation frequencies the antenna are electrically short, i.e. much shorter than the half wavelength, and the field of view  (FOV) extends almost the whole sky, the 2-dimensional plan-parallel approximation commonly used in small FOV interferometry imaging synthesis is no longer valid. Even the traditional large-FOV imaging algorithms developed for the ground-based arrays are not applicable either, as the array has moving baselines which has a three-dimensional distribution. Furthermore, for each baseline, the Moon blocks a slightly different part of the sky, making the synthesis even more complicated. 

\begin{table}
	\centering
	\caption{Basic parameters of the DSL lunar orbit array.  }
	\label{tab:basic parameters}
	\begin{threeparttable} 
	\begin{tabular}{lc} 
		\hline
		Array Design  & Values               \\
		\hline
		Number of interferometry satellites  & 5 -- 8 \\
		Number of antenna polarizations & 3\\
        Maximum baseline  & 100 km\\
        Minimum baseline  & 1 km\\
		Orbit height & 300.0 km\\
		Orbital plane inclination & $30.0^\circ$   \\
		Precession period   & 1.3 year    \\
		Total observation time  & 3 -- 5 years \\
		Frequency range & 1 -- 30 $\MHz$  \\
		Channel bandwidth & 8 kHz \\
		Number of frequency channels & 30\\
        \makecell[l]{Equivalent receiver input noise voltage} & $3\mr{nV/\sqrt{Hz}}$\tnote{**}      \\
		\hline
	\end{tabular}
	\begin{tablenotes}
	\footnotesize
	\item[**] The receiver noise is discussed in \S \ref{subsec:noise}. 
	\end{tablenotes}
	\end{threeparttable}
\end{table}

Despite of these complications, as long as the interferometric visibilities are linearly related
to the sky temperature, and this linear relation is precisely known, in principle it can be inverted mathematically, and a general formalism of lunar orbiting array imaging synthesis has been developed \citep{2018AJ....156...43H}. This approach can accommodate various complications faced by the DSL mission, including the all-sky FOV, 3-D moving baselines, the problem of mirror symmetry with respect to a single plane of baselines, and the varying sky blockage by the Moon. The performance of the basic algorithm with random baselines has been demonstrated for the noise-free case. Here we apply this map-making algorithm to the specific DSL mission, and take into account more practical issues, such as the thermal noise, the 3-D baselines generated by orbital motion of the satellites, the antenna response, and the varying Moon blockage, etc. 

To assess the scientific capability of such an array on lunar orbit, we run a set of imaging simulations, and investigate the imaging quality and the point source sensitivities of the array in this paper. Our simulation includes three steps: 1) construct a high-resolution sky map from a low-frequency sky model; 2) generate mock visibilities accounting for various practical issues expected for a lunar orbiting array; 3) reconstruct the sky image from the simulated visibility.

The present work serves as a basic version of the end-to-end simulation. There are other issues that will affect the imaging results, such as the measurement errors on the baselines vector and  time synchronization, the calibration errors on the magnitude and phase of the instrumental gain,  the slight deviation of satellites away from the same orbit, the RFI from the satellites themselves, etc. Those effects are more dependent on the specific instrument design and correction techniques, and we shall consider them in subsequent studies. In the present work, we shall simulate the basic performance of a DSL-like lunar orbiting array, and check how this depends on the array configuration. 

This paper is organized as follows. In Sec. \ref{sec:model}, we first introduce the model set up, including the input sky model, noise model, and orbit parameters. Then in Sec.\ref{sec:Algorithm}, we introduce the algorithm of map-making, and describe our treatment of the practical issues when generating the mock visibilities. 
In section \ref{sec:results}, we show our simulation results for the image reconstruction, including the all-sky imaging quality and the point source sensitivities, and discuss the effects of different design parameters. 
We discuss our results and conclude in section \ref{sec:discuss}.

Throughout the paper, we use boldface letters to denote data vectors and matrices. For a matrix $\mathbf{A}$, $\mathbf{A}^H$ denotes its Hermitian conjugate, $\mathbf{A}^{-1}$ denotes its inverse, and $\mathbf{A}^\dagger$ denotes its Penrose-Moore pseudo-inverse.

\section{Model}
\label{sec:model}

In the current design, the DSL array consists a mother satellite and a number of daughter satellites orbiting the Moon along the same orbit, and a linear interferometer array is formed by the daughter satellites equipped with antennas. Interferometric observations are made by the daughter satellites, preferably in the part of orbit where the Earth is shielded. 
The basic parameters of the DSL concept that are relevant to our imaging simulation are listed in Table \ref{tab:basic parameters}.

In our imaging simulation, we first generate time-ordered interferometric visibility data by assuming an input sky map and a beam model,  taking into account the orbital motion, the Moon blockage, the antenna response, and with an additive random noise.   
This mock data is then processed to reconstruct the sky map by applying the generic imaging algorithm developed by \citet{2018AJ....156...43H}. 
In this section we shall describe our model setup, while the imaging algorithm and array configurations are discussed in the next section.

\subsection{Input sky map}
\label{subsec:skymodel}

\begin{figure}
	\centering
    	\label{fig:subfig:10MHz}
		\includegraphics[width=0.8\columnwidth]{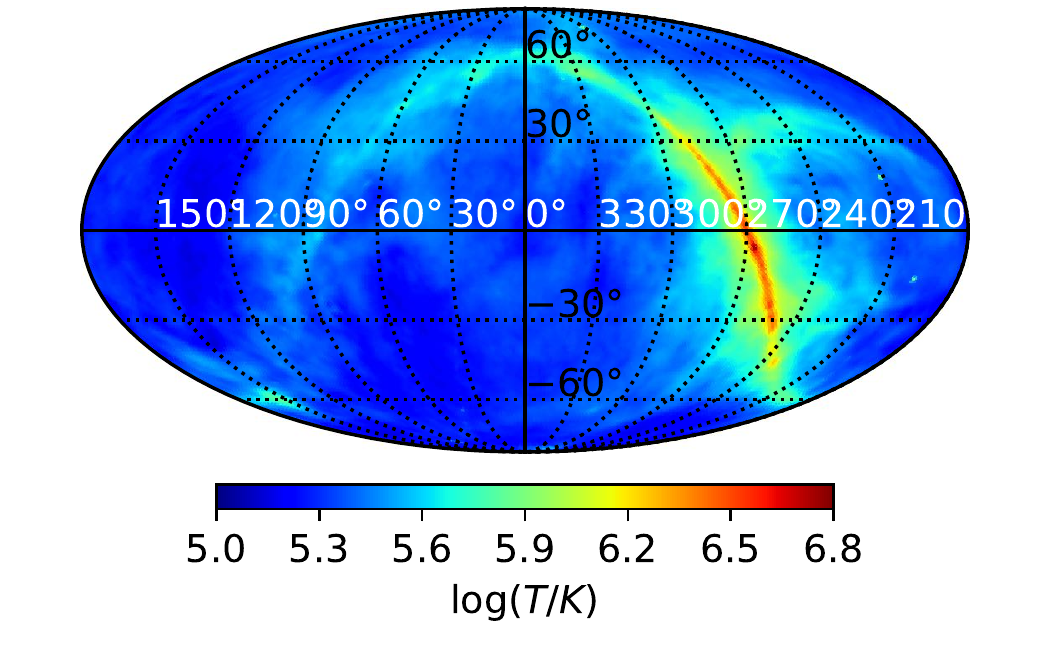} 
    \caption{
    The whole sky map at 10 MHz in the ecliptic coordinate system.
    }
    \label{fig:input_map}
\end{figure}

At present, full sky maps in the frequency range of 1 - 30 MHz are not available. We therefore 
use a sky map generated by extrapolation from higher frequencies. A number of extrapolated sky models have been developed, such as the Global Sky Model (GSM) \citep{2008MNRAS.388..247D,2017MNRAS.464.3486Z}, and the Cosmology in the Radio Band (CORA) \citep{2014ApJ...781...57S,2015PhRvD..91h3514S}.  In the present work,  we make use of the high-resolution Self-consistent whole Sky Model (SSM)  \citep{2019SCPMA..6289511H}.
Note that the SSM model used by the current work has not accounted for the absorption effect that would become very significant at $\nu < 10 \MHz$. A model which incorporates the free-free absorption effect has been developed by \citet{2021arXiv210403170C}.
For the purpose of evaluating the map-making capability, however, the original SSM model is fairly adequate, and maybe even more preferable as it gives a familiar sky map that is easier to compare at different wavelengths. This model may yield a slightly higher sky brightness temperature than the actual one, as it ignores the interstellar absorption.

In the full-sky simulation below, we pixelize the map with a resolution of about $1^\circ$, corresponding to HEALPIX pixels with $N_{\rm side}=64$. 
The input sky map at 10 MHz is shown in Fig.~\ref{fig:input_map}.

\subsection{Orbit parameters}
\label{sec:orbit}
We shall consider a nearly circular orbit for the array of satellites. In choosing the orbit parameters, we keep in mind three aspects of the problem: (1) the stability of the orbit; (2) the good observation time during which the Earth (and the Sun) is shielded. (3) the nodal precession of the orbit which generates a three dimensional distribution of baselines, which is important for breaking the mirror symmetry in image synthesis \citep{2018AJ....156...43H}. Some of these may conflict with each other, for example,  a lower orbit will allow a larger part of the sky being blocked, thus increasing the good observation time, but the lower orbit is also less stable, which would require more orbit maintenance to ensure the safety of the array.  We have to weigh these factors and reach a compromise in our choice.

\begin{figure}
	\centering
	\includegraphics[width=1.0\columnwidth]{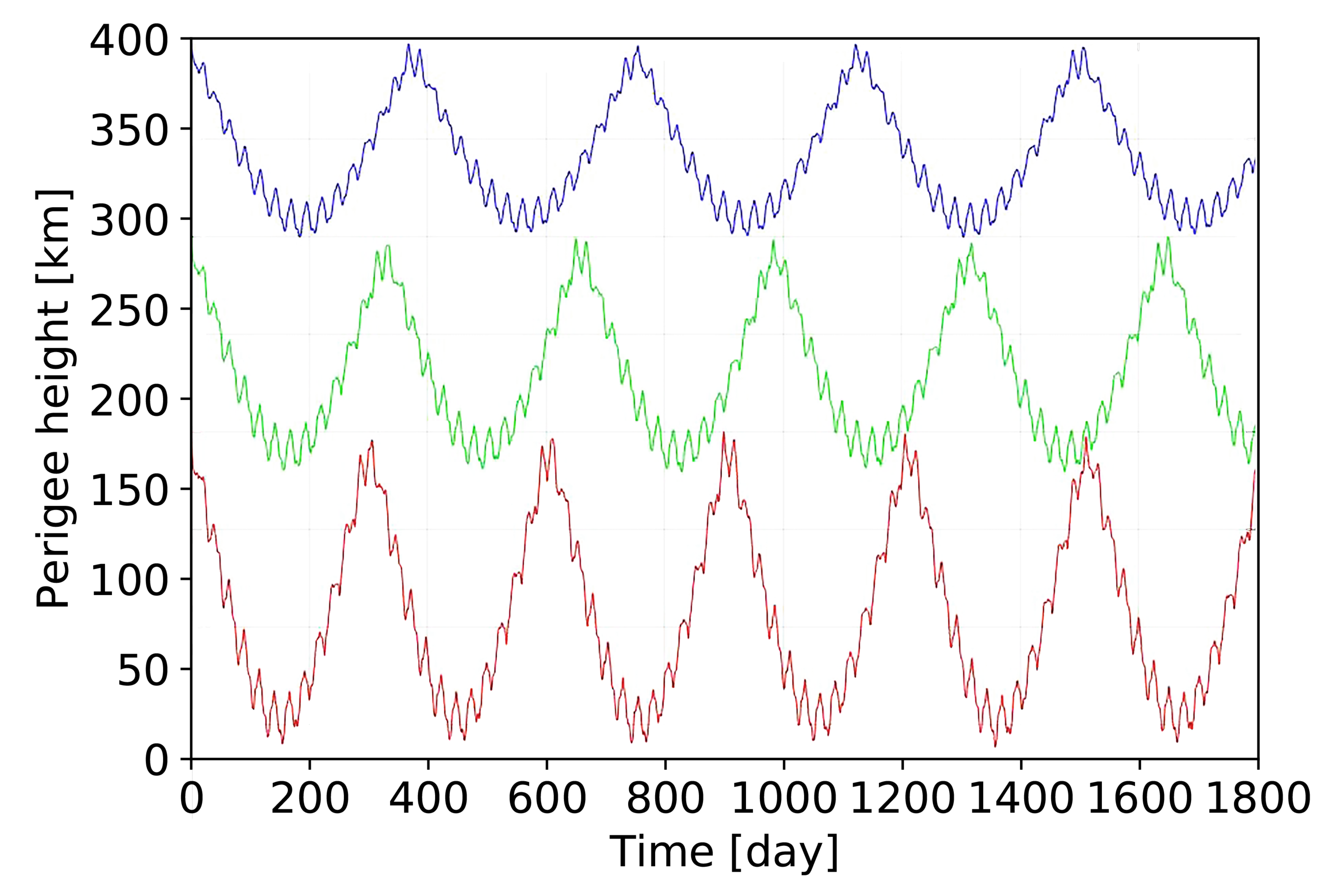}
    \caption{Variation of the perigee height in 5 years. The three curves correspond to three orbits with the average height of 400 km, 300 km, and 200 km, from top to bottom respectively. }
\label{fig:height}
\end{figure}

\begin{figure}
	\centering
	\includegraphics[width=0.99\columnwidth]{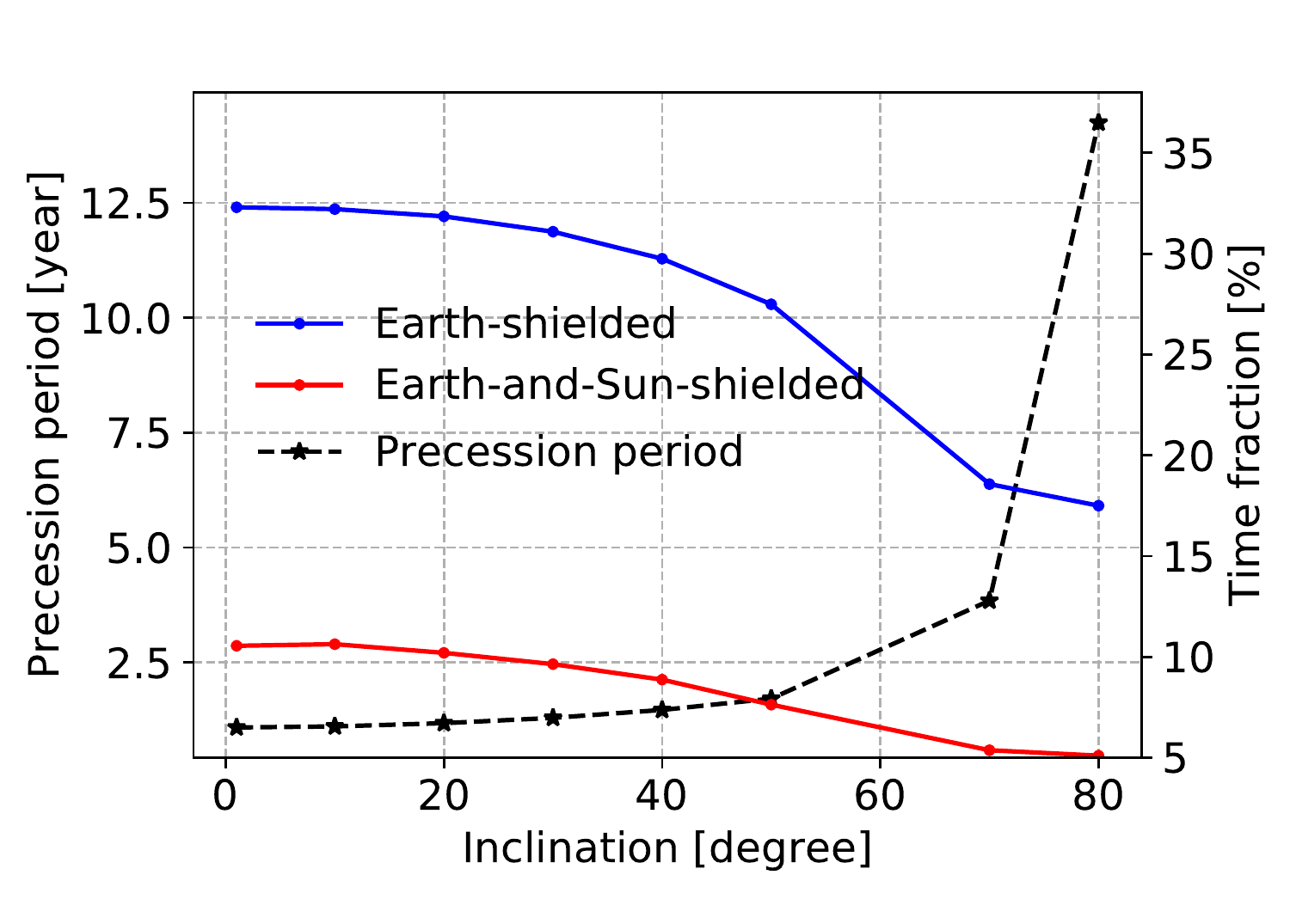}
    \caption{The dashed line (reading the numbers from the {\it left axis}) shows the orbit precession period for different inclination angles. The solid lines (reading the numbers from the {\it right axis}) correspond to the fraction of good observation time as a function of inclination angle. The blue curve shows the time fraction when the Earth is shielded, and the red curve shows the fraction of time when both the Earth and the Sun are shielded.}
    \label{fig:shield}
\end{figure}

In order to evaluate the merit of each orbit choice, we carry out orbit simulations. 
We select the GL1500E lunar gravity field model \citep{GL1500E}, in which the order of the model is $100\times100$, and the JPL DE405 Planetary ephemeris\footnote{\url{https://ssd.jpl.nasa.gov/?planet_eph_export}} is used.
 In Fig.~\ref{fig:height}, we plot the perigee height of three orbits of different average height over a period of 5 years. We see that as the orbit precesses, there is a variation in its perigee height. For the orbit of 300 km average height, the minimum height of the perigee is 180 km during the the procession cycle, while for the 200 km average, the  the minimum height of the perigee is 38 km. On the other hand, the mean half angle extended by the Moon is given by 
\begin{eqnarray}
\sin\theta = \frac{R}{R+h}
\end{eqnarray}
where $R=1737.4 $km is the mean radius of the Moon, so $\theta= 58.51^\circ$ and $63.74^\circ$ for the 300 km and 200 km respectively. 
We have tentatively chosen a 300 km as the fiducial value of the orbit height for the project.

The Moon revolves around the Earth in an orbit which is only slightly inclined ($\sim 5^\circ$) with respect to the ecliptic plane, and the
lunar equator is only $1.543^\circ$  inclined w.r.t. the ecliptic plane, so we can approximate both with the ecliptic. Due to orbital dynamics, 
the orbital plane of the lunar satellite array will precess. This is very important for the imaging synthesis, because for the linear array the baselines are all distributed on the orbit plane, without precession the orientation of this plane will be fixed and there is a mirror symmetry with respect to this plane. The precession allows the plane to rotate and orient toward different directions, resulting in a three dimensional distribution of baselines, which is essential to break the mirror symmetry. Shorter orbit precession cycle allows faster filling of the three dimensional $uvw$ space. In Fig. \ref{fig:shield}, we plot the precession period (reading number from the left axis), for different  orbital inclination
angles for the 300 km orbit. The precession rate decreases with the inclination angle, the 
cycle is longer than 2 years for orbits with inclination angle $>55^\circ$, and unsafe for $60^\circ$. On the other hand, although the precession is faster at lower 
inclination angles, the resulting baseline distribution would be compressed into a disc in the $uvw$ space if the inclination angle is too small, limiting the angular resolution in the direction perpendicular to the ecliptic plane. 

Another factor to be considered is the percentage of the good observation time, i.e. the time when the Earth and/or the Sun is shielded by the Moon from the view of the satellites.
Here we take a geometric 
optics definition of shielding, though due to diffraction some RFI may still be observed at the edge of the shielded region. 
The time when the Sun and the Earth are both shielded is the intersection of them.
This fraction is maximized at zero inclination, and decreases with increasing declination. This is because at zero inclination angle, the Earth, the Moon and the satellites are all on the same plane, so there is always time when the satellites are shielded when they orbit the Moon, but for large inclination angle the satellites will spend some time outside the plane of the Moon's orbit, during which they are more easily exposed. In Fig. \ref{fig:shield}, with the right axis, we plot the fraction of good observation time for a 300 km orbit with different inclination angles during five years, and we also plot the ``double good'' time when both the Sun and Earth are shielded.

Based on these considerations, we choose a 300 km circular orbit with an inclination angle of $30^\circ$ as our 
fiducial orbit. For this orbit, the precession cycle period is about 1.3 year, the good observation time that the Earth is shaded is 
about $31\%$,  and the ``doubly good'' observation time both Earth and Sun are shielded is about
$10\%$.

\subsection{System Noise}
\label{subsec:noise}

The thermal noise on the measured visibility in temperature units is  
\begin{equation}
	\sigma_n=\frac{T_{\mr{sys}} }{\sqrt{\Delta \nu t_{\mr{int}}}},
	\label{eq:noise}
\end{equation}
where $\Delta \nu$ is the bandwidth of the channel,  and $t_{\text{int}}$ is the integration time. Based on the preliminary design of the DSL, the basic
channel width is set at  $\Delta \nu = 8 \kHz$,  but note that multiple channels may be combined to increase the sensitivity.

\begin{figure}
	\centering
	\includegraphics[width=1\columnwidth]{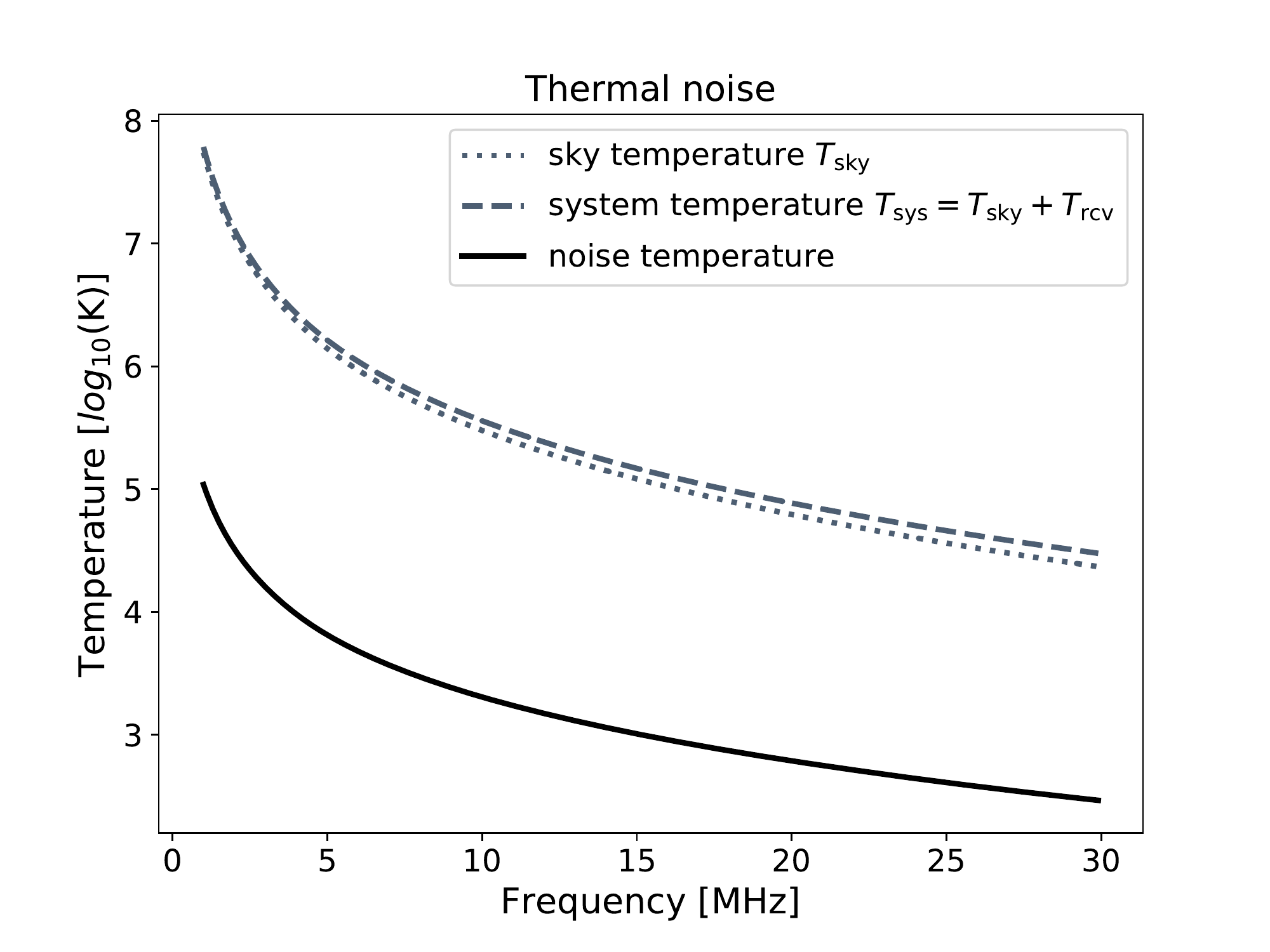}
  	\caption{The sky brightness temperature (dotted line), the total system temperature (dashed line), and the noise temperature (solid line).
  	The noise temperature is evaluated by Eq.(\ref{eq:noise}) with the longest integration time determined by Eq.(\ref{eq:int_time}) for the 1 km baseline, i.e. 26.26 s for 1.5 MHz, and 13.13 s for 3 MHz, etc. 
  	Note the system temperature is always dominated by the sky brightness temperature 
  	at $1 - 30$ MHz. }
	\label{fig:noise}
\end{figure}

The system temperature $T_{\mr{sys}}$ receives contributions from both sky radiation and receiver noise,
\begin{equation}
T_{\rm{sys}} = T_{\rm{rcv}} + T_{\rm{sky}}.
\end{equation}
Here $T_{\rm{rcv}}$ is the effective temperature of receiver noise, and $T_{\rm sky}$ is the mean brightness temperature of the sky, as average over the beam of the antenna. In the Rayleigh-Jeans limit, the total energy flux received by a dipole antenna is given by
\begin{eqnarray}
S_\nu &=& \frac{1}{2} \int \frac{2 \nu^2 k_B T_{\rm sky}}{c^2}\, A(\theta,\phi)\,{\rm d}\Omega  \nonumber \\
&=& \frac{k_B T_{\rm sky}}{\lambda^2}\,\frac{4\pi}{G},\label{eq.Snu}
\end{eqnarray}
where $A(\theta,\phi)$ is the antenna beam pattern, $G$ is the gain which equals to 1.5 for short dipoles, and the factor 1/2 in first equation is due to the fact that the antenna responds to only one 
polarization.
The design of the receiver for the DSL low frequency interferometry system will be presented elsewhere, but based on preliminary design, each polarization of the DSL low frequency band system is made up of a pair of 
monopole antenna. The signal from each antenna is individually amplified before combined and digitized. The receiver noise on each side is given in terms of equivalent voltage fluctuation at its input stage, with
$\approx 3\, {\rm nV}/\sqrt{\rm Hz}$ 
over the observation band. As the noise on the two sides are uncorrelated, the combined noise is 
\begin{equation}
V_{\rm n} \approx 3 \sqrt{2} \, {\rm nV}/\sqrt{\rm Hz}.
\end{equation}
This is to be compared with the voltage induced in the antenna. 
The equivalent energy flux of the receiver noise, that is to be combined with the flux from the sky (Eq.(\ref{eq.Snu})),
is given by
\begin{equation}
S_{\rm n} =   \frac{1}{Z_0}\left(\frac{V_{\rm n}}{l_{\rm eff}}\right)^2,
\end{equation}
where $Z_0 = 377\, \Omega$ is the impedance of the vacuum, and $l_{\rm eff}$ is the effective length of the dipoles. 
Then the effective receiver temperature is
\begin{equation}
T_{\rm rcv} = \frac{G\, V_{\rm n}^2}{4\pi\,k_B\, Z_0\,\left(\frac{l_{\rm eff}}{\lambda} \right)^2}.
\end{equation}
In this work, we assume the length of the dipoles is 5 m, corresponding to $l_{\rm eff}\approx 2.5 \rm m$.
Thus, the receiver noise is equivalent to 
\begin{equation}
S_{\rm n} = 7.64 \times 10^{-21}\, {\rm W/m^2/Hz},
\end{equation}
or
\begin{equation}
T_{\rm rcv} = 6.61  \times 10^3 \left(\frac{\lambda}{10 \m}\right)^2 \left(\frac{2.5 \m}{l_{\rm eff}}\right)^2 \K.
\end{equation}
These are  plotted in  Fig.~\ref{fig:noise}. This receiver temperature increases as $\lambda^2$, while the Galactic synchrotron radiation, in the optical thin regime, scales as $\lambda^{2.6}$, until at very low frequency when absorption becomes important. Hence at the low frequency band considered here, the sky radiation dominates the system temperature even for the relatively short antennas of DSL.

The satellites will orbit the Moon with a period of $\sim 2.3$ hour. 
The longer baselines are moving fairly fast, and the integration time for a visibility measurement is limited to the baseline crossing time.
For a given observing frequency and a specific baseline, the change in the projected baseline length should be less than $\sim 1/10$ of the wavelength within the integration time, in order to avoid the smearing in the $(u,v,w)$ sampling.  Therefore, the integration time of any visibility measurement is restricted to
 \begin{eqnarray}
 	t_{\mr{int}} \leq \frac{2}{\omega} \arcsin \left(\frac{1}{20} \frac{\lambda}{D}\right)
 	\label{eq:int_time}
 \end{eqnarray}
where $\omega$ is the angular velocity of the satellites, and $D$ is baseline length. 
For the DSL orbit, the maximum integration time is 0.013 seconds for the 100 km baseline at 30 MHz.

The sky temperature and noise temperature at different frequencies are shown by the dotted and solid lines respectively in Fig.~\ref{fig:noise}.  Note that although the longer baselines have much higher noise due to the shorter integration time, 
they have much more measured visibility data points.

\section{Algorithm}
\label{sec:Algorithm}

In the following, we choose 1.5 MHz and 10 MHz to show results from the full-sky simulations,  
and choose several frequencies in presenting the results from part-sky simulations.

\subsection{Imaging algorithm}
\label{subsec:algorithm}

The visibility data $V_{ij}$ measured by an interferometer array elements pair $i, j$ is related 
to the sky brightness $I(\bm{\hat{k}})$ by
\begin{equation}
	V_{ij} = \int A_{ij} (\bm{\hat{k}}) I(\bm{\hat{k}})~ e^{-i \qvec{k} \cdot \qvec{r}_{ij}}  d^2 \bm{\hat{k}},
	\label{eq:Vij}
\end{equation}
where $A_{ij}(\bm{\hat{k}})$ is the combined antenna beam pattern that depends on the antenna primary beam response, 
$I(\bm{\hat{k}})$ is the sky brightness in the direction $\bm{\hat{k}}$, $\qvec{k}=\frac{2 \pi }{\lambda} \bm{\hat{k}}$ is the wave vector, 
and $\qvec{r}_{ij}=\qvec{r}_i-\qvec{r}_j$ is the baseline vector formed by antennas $i$ and $j$.
The visibility function can be re-written in the $uvw$ coordinates given in wavelength units $\qvec{u}= (u,v,w) \equiv \qvec{r}/\lambda$,
\begin{eqnarray}
	V(u,v,w) = \int \frac{{\mr d}l\, {\mr d}m}{n} A_{ij}(l,m)  I(l,m)  e^{-i2\pi\left[ ul + vm + w(n-1) \right]}.
\label{eq:Vuvw}
\end{eqnarray}
where $(l, m, n)$ are the direction cosines with respect to the three coordinate axes, and $l^2+m^2+n^2=1$.

For ground-based arrays with small FOV, the $w$-term in the exponential can often be neglected, such that the integral in Eq.(\ref{eq:Vuvw}) can be reduced to a 2D Fourier transform, and with the beam $A(l,m)$ known, the sky intensity can be easily recovered by inverse Fourier transform. For wider FOV, the image can still be recovered by a 3D Fourier transform, and also a number of algorithms were developed for such wide-field imaging \citep{2017isra.book.....T}. However, for the case we are considering here, these algorithms are not suitable for several reasons. First, there is no ground to shield half of the sky, and by using electrically short dipole antennas, nearly the whole sky is within the FOV. For data taken during one orbit, where the baselines are distributed on a circular ring on the orbital plane, there is a mirror symmetry problem: a source located on one side of the plane produce exactly the same visibility as a source located at the mirror image position on the other side of the plane, the two can not be distinguished \citep{2018AJ....156...43H}. Second, the precession of the orbital plane produces a three-dimensional distribution of baselines. While this breaks the mirror symmetry, the $w$-term could never be neglected. Third, the Moon will block a fraction of the sky, but at each location on the orbit, the blocked part is different.  Strictly speaking, the visibility data at each time corresponds to a sky with a different blockage, so even the 3D Fourier transform method is not applicable. 

However, if we neglect the reflection and radiation of the Moon, the visibility is still linearly related to the sky intensity, so in principle the sky intensity can be recovered by solving this time-dependent linear equation
\citep{2018AJ....156...43H}, which can work in spite of all the issues listed above.
The Moon blockage can be introduced as a shade function into the visibility, i.e.
\begin{eqnarray}
	V_{ij} = \int A_{ij} (\bm{\hat{k}}) S_{ij}(\bm{\hat{k}}) I(\bm{\hat{k}})~ e^{-i \qvec{k} \cdot \qvec{r}_{ij}}  d^2 \bm{\hat{k}},
	\label{eq:Vij-2}
\end{eqnarray}
where $S_{ij}$ is the shade function that describes the time-dependent positions in the sky blocked by the Moon. We pixelize the sky with the HEALPix scheme \citep{2005ApJ...622..759G}, and the integral over sky angles is discretized as a sum over sky pixels, 
\begin{eqnarray}
	V_{ij}(t) = \sum_{\alpha=1}^{N_{\rm pix}} B(\alpha,t)\, I(\alpha) \Delta\Omega
    \label{eq:Vijn}
\end{eqnarray}
where $\alpha$ is the pixel index, $N_{\rm pix}$ is the total number of pixels,
$\Delta\Omega=4\pi/N_{\rm pix}$ is the solid angle corresponding to one pixel, and $B(\alpha,t)$ is the complex response of the array, which is an effective ``beam'' taking into account of both antenna beam response and the Moon blockage, i.e.
\begin{eqnarray}
B(\alpha,t)=A_{ij}(\alpha,t)\, S(\alpha,t) e^{-i \qvec{k}_{\alpha} \cdot \qvec{r}_{ij}(t)}.
	\label{eq:B}
\end{eqnarray}

Converting the sky brightness $I$ to the brightness temperature $T$ according to the Rayleigh-Jeans approximation, and considering the thermal noise, the observed
visibilities can be written in a matrix form as
\begin{eqnarray}
\mathbf{V} = \mathbf{B T + n}.
\label{eq:matrixV}
\end{eqnarray}
With a total of $N_{\mr{t}}$ observation time points and $N_{\mr{bl}}$ baselines, $\mathbf{B}$ is a $(N_{\mr{t}} \cdot N_{\mr{bl}})\times N_{\mr{pix}}$ matrix, $\mathbf{T}$ is a $N_{\mr{pix}}$-element vector with each element corresponding to one pixel in the sky, and the visibility $\mathbf{V}$ and noise $\mathbf{n}$ have the dimension of $(N_{\mr{t}} \cdot N_{\mr{bl}})$. 
In the following, we adopt the convention  $\sum_{\alpha=1}^{N_{\rm pix}} A_{ij}(\alpha)\, \Delta\Omega = 1$, so that the visibility has the same units as the sky brightness temperature.

We model the noise as a Gaussian random variable with zero mean, i.e. $<n>=0$, variance $\sigma_n^2$ (given by Eq.~(\ref{eq:noise})), and the noise covariance matrix $\mathbf{N} \equiv \langle \mathbf{n n}^\mr{H} \rangle$.
The minimum variance estimator (for Gaussian noise, also the maximum likelihood estimator) of $\mathbf{T}$ is then 
\begin{eqnarray}
\mathbf{\hat{T}=(B^\mr{H} N^{-1} B)^{-1} B^\mr{H} N^{-1} V. }
	\label{eq:T1}
\end{eqnarray}
Note that $\mathbf{B^\mr{H} N^{-1} B}$ is a $N_{\mr{pix}} \times N_{\mr{pix}}$ matrix.
However, if $\mathbf{ B^\mr{H} N^{-1} B}$ is not invertible, which is often the case due to limited sampling in the $uvw$ plane, we can still compute some psuedo-inverse matrix, e.g. Moore-Penrose pseudo-inverse \citep{2018AJ....156...43H}: 
\begin{eqnarray}
\mathbf{D} \sim \mathbf{ (B^\mr{H} N^{-1} B)^{\dagger}}.
\end{eqnarray}

More generally other estimators of $\mathbf{T}$ can be formed with different choices of $\mathbf{D}$ \citep{2015PhRvD..91b3002D}.
The imaging quality can be quantified by the point spread function (PSF) of the reconstruction, and the covariance of the estimator.
The expectation of the above estimator is 
\begin{eqnarray}
\langle\hat{\mathbf{T}}\rangle &=& \langle \mathbf{D B^\mr{H} N^{-1} V} \rangle \nonumber\\
&=& \mathbf{D (B^\mr{H} N^{-1} B) T } \nonumber\\
&\equiv& \mathbf{P T},
\label{eq:mean}
\end{eqnarray}
where 
\begin{eqnarray}
\mathbf{P = D (B^\mr{H} N^{-1} B)}
\label{eq:psf}
\end{eqnarray}
is the matrix-valued point spread function. If the matrix  $\mathbf{B^\mr{H} N^{-1} B}$ is invertible  we would have the idealized PSF $\mathbf{P}=\mathbf{I}$. 
Note that $\mathbf{B^\mr{H} N^{-1} B}$ fully
characterizes the observation, and  
may be regarded as the dirty beam of the system. If we simply set $\mathbf{D}$ as a normalization matrix in Eqs.~(\ref{eq:mean})-(\ref{eq:psf}), then we obtain the so called ``dirty image'' without deconvolution. This PSF of the dirty beam shows the basic characteristics of the specific array configuration.

The covariance of the estimator is 
\begin{eqnarray}
\mathbf{C}
 &\equiv& \mathbf{\left\langle(\hat{T}-\langle\hat{T}\rangle)(\hat{T}-\langle\hat{T}\rangle)^{\mr{H}}\right\rangle = PD^{\mr{H}}
     }
\end{eqnarray}
In the case of sufficient $uvw$ sampling with an invertible $\mathbf{B^{\mr{H}}N^{-1} B}$ matrix, the covariance 
$\mathbf{C\to (B^{\mr{H}}N^{-1} B)^{-1}}$. 
In our Moore-Penrose pseudo-inverse approach, the matrix $\mathbf{B^{\mr{H}}N}^{-1} \mathbf{B}$ is approximately the inverse of the covariance, which measures the information content of the reconstructed map.
In Section \ref{sec:psf} we will calculate and analyze the PSF and eigenvalue spectra of the $\mathbf{B^{\mr{H}}N}^{-1} \mathbf{B}$ matrix for our different array configurations.

\subsection{Doppler Effect}
In the above, we have assumed the satellites are at rest with each other. Indeed, for an array of satellites moving on the same circular
orbit and located close to each other, the velocity differences are small and this is a good approximation. However, we will make sky map by synthesizing the data collected over a period of several years, during this time, the orbital motion of the satellites, Moon, and Earth induce varying
velocities with respect to the inertial frame, so here the effects induced by the motion need to be considered.

In the presence of velocity, up to the first order,  
$\nu' = \nu (1- \bm{\beta} \cdot \bm{\hat{k}})$, and $\qvec{k}' = \qvec{k}/(1+\bm{\beta}\cdot \bm{\hat{k}})$, 
where $\bm{\beta} = \qvec{v}/c$. 
The visibility is given by 
\begin{eqnarray}
V_{\text{ab}}(t,\nu) &=& \int d^2\bm{\hat{k}} B_{\text{ab}} (\nu, \bm{\hat{k}}) I(\nu', \qvec{k}') e^{-i \qvec{k}' \cdot \qvec{r}_{\text{ab}}}\\
&\approx &\int d^2\bm{\hat{k}} B_{\text{ab}}(\nu, \bm{\hat{k}}) I[\nu (1- \bm{\beta} \cdot \bm{\hat{k}}), \qvec{k}/(1+\bm{\beta}\cdot \bm{\hat{k}})] \nonumber\\ 
&\times & e^{-i(1-\bm{\beta}\cdot\bm{\hat{k}})(\qvec{k} \cdot \qvec{r}_{\text{ab}})}.\nonumber 
\end{eqnarray}

Simple estimates show that the circular motion around the Sun has a magnitude of  $v \sim 30 \kms$, while the lunar  motion around the Earth ($v \sim 1 \kms  $) and the satellite motion around the Moon ($v \sim 1.7 \kms$) are smaller.  
For the heliocentric motion which has the largest effect,  $\beta \sim 10^{-4}$. For the continuum radiation, this induces a correction which is much 
smaller than the current precision of the measurement, so at first approximation, we can take $I(\nu', \qvec{k}') \approx I(\nu, \qvec{k})$. However,  
the correction on the phase is not negligible, so we have  
\begin{eqnarray}
V_{\text{ab}}(t,\nu) &\approx &\int d^2\bm{\hat{k}} B(\nu, \bm{\hat{k}}) I[\nu, \bm{\hat{k}}] e^{-i(1-\bm{\beta}\cdot\bm{\hat{k}})(\qvec{k} \cdot \qvec{r}_{\text{ab}})}
\end{eqnarray}
The correction  $(\bm{\beta}\cdot\bm{\hat{k}})(\qvec{k} \cdot \qvec{r}_{\text{ab}}) \sim \mathcal{O}(1)$ for the longest baseline of 100 km at wavelength of 10 m. Note that in ground-based
observation, this effect also exists, but it is usually corrected in real time when making the observation as the field of view is usually small and this phase factor can be directly compensated for a given direction. In our case, the visibility receives contribution from the whole sky, so the correction can not
be directly applied to the data. However, this phase is also deterministic, and one can include this phase factor in the mapping matrix $\mathbf{B}$, and correct for the velocity effect in the map reconstruction procedure.

Another possible problem is that if there is a spectral line from an object in a certain direction, and if we make imaging observation using 
narrow channels, i.e. the channel width is smaller than or comparable with the line width, then the line may be shifted outside the channel. However, if the source of such a line emission is known, one can make the observation by making real-time phase correction on the data for
the given direction, just as is used in ground-based observations. In principle, one can also apply such a correction for every sky direction to obtain the whole sky map for the narrow spectral line, though that requires much more computation.

\subsection{Array configurations}
\label{subsec:beam}

\begin{figure}
	\centering
	\includegraphics[width=0.9\columnwidth]{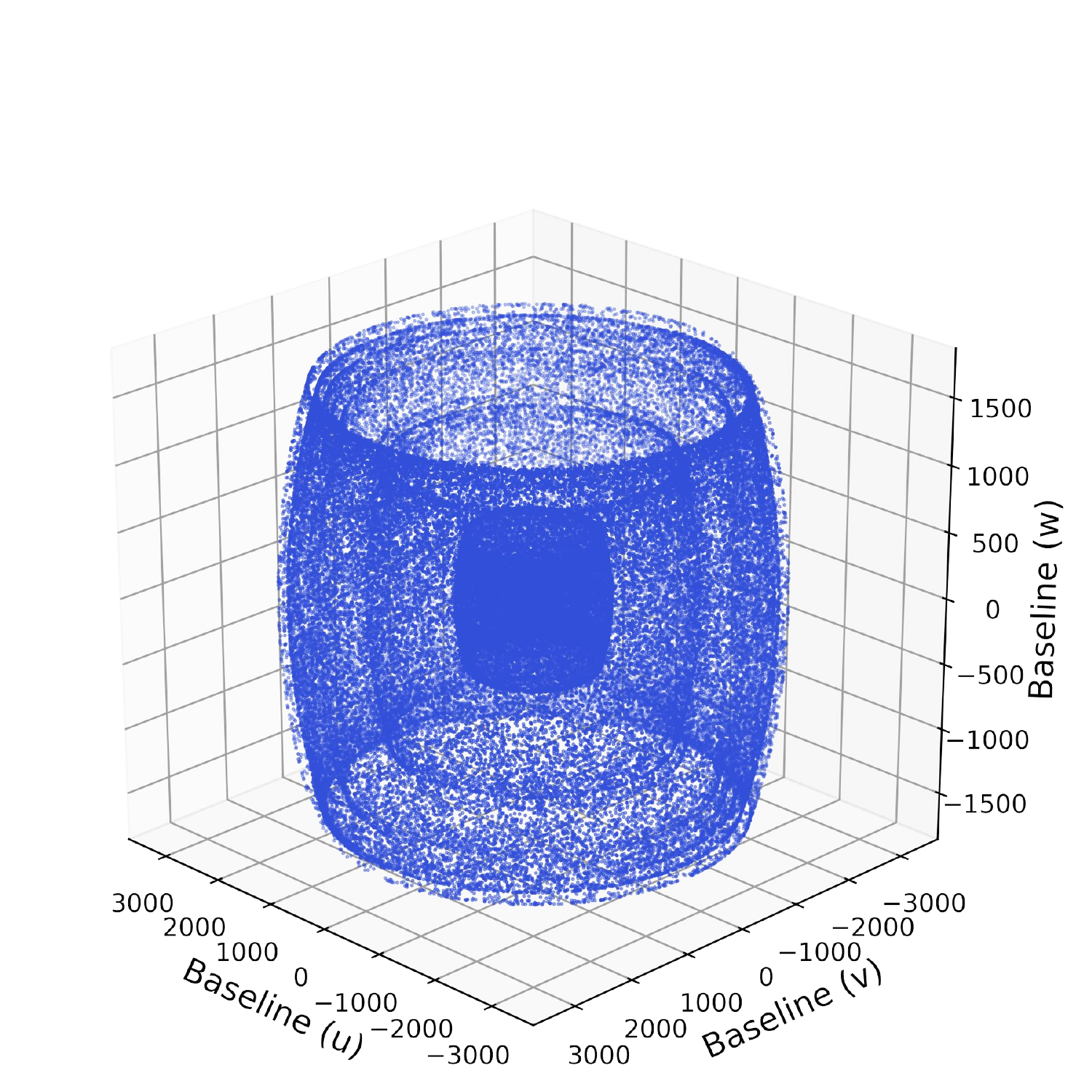}
	\caption{The three-dimensional baseline distribution of 5 satellites with optimized distribution. To illustrate the distribution, 
$10^5$ randomly selected points out of  $5.3\times 10^8$ observational points in the $uvw$ space are plotted.}
	\label{fig:baseline_5s}	
\end{figure}

With the electrically short antennas which are omnidirectional, and without a ground shielding one-half of the sky, there is a mirror-symmetry problem \citep{2018AJ....156...43H}.
However, because of the nodal precession of the orbit, we can naturally acquire a three dimensional (3-D) distribution 
of baselines to break this mirror-symmetry. As an example, the baselines generated by 5 satellites are 
shown in Fig.~\ref{fig:baseline_5s}, which has an onion-like layered structure.

The number of the daughter satellites considered here is in the range of $5 \sim 8$, to be determined by the mass of the payload and the capability of the launching rocket. After the number of satellites is decided, there are also different options for the array configuration, i.e. the relative positions and baseline lengths of the satellites. 
As a start, we set the minimum distance between any of the daughter satellites to be fixed as 1 km, which provides an ample safety margin  against satellite collision. The maximum distance is set to be 100 km, within which precise angular position measurement can be achieved straightforwardly with small star sensors, and at the same time it also provides an angular resolution which approaches the static imaging resolution limit imposed by interplanetary and interstellar scintillation \citep{2009NewAR..53....1J}. 

We consider four configurations of the linear array as follows, and investigate the imaging quality in each case.

{\bf a. Even spacing. }
In this case the daughter satellites are equally spaced, except for the spacing between the first and second satellites which forms the shortest baseline of 1 km. The distance between the $i$-th satellite and the first satellite is
\begin{equation}
 \begin{cases}
L_{2} &= 1 \km,  \\
L_{i-1} &=1+\frac{99}{N-2}(i-2) \km,  3\le i<N,\\
\end{cases}
\label{eq:even}
\end{equation}
where the $N$ is total satellite number.

{\bf b. Random spacing.}
In this case, after fixing the first two satellites and the last one to ensure the minimum and the maximum baselines, we randomly insert the other satellites between the second and the last satellite with the minimum distance between any two satellites being larger than 1 km. One such realization is used, which we refer to as the random spacing\footnote{The position of the 8 satellites are set at 0 km, 1.0 km, 2.89 km, 10.17 km,  24.23 km,  43.23 km, 69.75 km, and 100.00 km with respect to the first satellite, respectively. }, though strictly speaking the word ``random'' should refer to a distribution of baselines, not a single case. 

{\bf c. Logarithmic spacing.}
We also test the logarithmic spacing  between the satellites, in which the position of the $i$-th daughter satellite is set at
\begin{eqnarray}
	L_i = 10^{\frac{2}{N-2} (i-2)} \km, \qquad 2\le i\le N.
\end{eqnarray}
The basic idea here is to produce baselines on all different scales.
 
{\bf d. Optimized spacing.}
The optimized spacing configuration is motivated by the consideration of achieving an $uvw$-coverage on all different scales as in the logarithmic case, but with an emphasize on having more long baselines. We propose a modified logarithmic spacing configuration, in which the
the $i$-th daughter satellite ($2<i<N$) is placed at a distance 
$L_{i}$ away from the first one, with
\begin{eqnarray}
L_{i} = 10^{2-\frac{1}{2^{i-3}}} \km, \qquad 3\le i<N,
\end{eqnarray}
and $L_2= 1\km$, $L_N= 100 \km$.

\begin{figure}
\centering
\includegraphics[width=0.9\columnwidth]{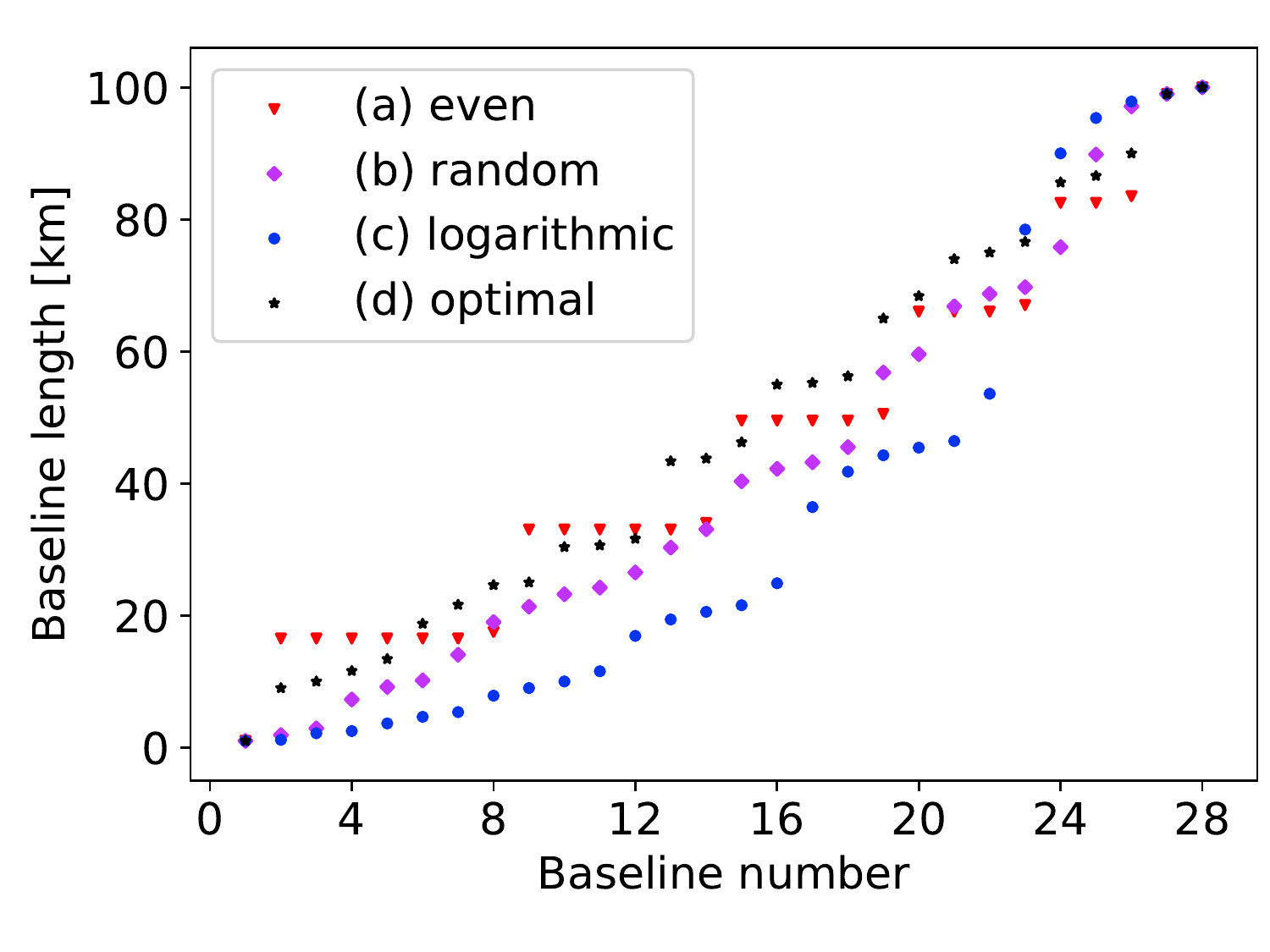}
\caption{The baseline length distribution of 8 satellites for the four configurations as indicated in the legend. In each case the 28 baselines range 
from 1 km to 100 km. Cases (a), (b), (c) and (d) are plotted with red triangles, purple diamonds, blue dots, and black stars, respectively. }
\label{fig:bl_length}	
\end{figure}

The length distribution of the 28 baselines formed by 8 satellites are shown in Fig.~\ref{fig:bl_length} for the above four configurations. While in all four 
cases the lengths are distributed across the whole range from 1 km to 100 km, the ``optimized'' (d) and ``random'' (b) spacing cases appear to spread out more smoothly. The even spacing (a) configuration has a high degree of redundancy, and the logarithmic spacing (c) strategy produce more short baselines.

In this work we shall model the
motion of the satellites as simple circular motion on a slowly rotating (precessing) orbital plane, and the satellites are fixed on the assigned place. The actual orbital motion of the satellites may be more complicated, as the orbit may not be a perfect circle, and is subject to the perturbation of the anomalous lunar gravitational field,
and each satellite may have slightly different orbital parameters. However, these would not change the overall imaging results, as long as the relative positions between the daughter satellites are measured with sufficient precision. The slightly different orbital parameters may actually allow the satellites to sample a slightly large part of $uvw$ space. Indeed, one may intentionally generate relative motions between the satellite, so that they periodically approach and then back away from each other. Such a ``breathing mode" would increase the $uvw$ space sampled over the operation period, but here we shall not consider these more complicated cases.

Given the array configuration, the coordinates of each satellite during the orbital motion is specified, and the varying baseline vector $\qvec{r}_{ij}(t)$ formed by each pair of the satellites can be calculated.
We shall compare the imaging results produced with the different array configurations  in section \ref{sec:results}.

\subsection{Beam}

\subsubsection{The Antenna Response}
\label{sec:response}
In the DSL mission concept, each imaging daughter satellite is equipped with three orthogonal pair of dipole antennas. A simple model of the short dipole can be used to approximate the antenna response:
\begin{eqnarray}
A(\beta)=\mr{sin^2}(\beta)
\end{eqnarray}
where $\beta$ is the angle between the observing direction and the direction of the antenna.
So the beam function formed by a pair of antennas is 
	\begin{equation}
	\begin{split}
	A_{ij}(\qvec{k},\qvec{R}_i) &= \sqrt{\mr{sin}^2(\langle \qvec{k},\qvec{R}_i\rangle)\times \mr{sin}^2(\langle \qvec{k},\qvec{R}_j\rangle)} \\
	&= |\mr{sin}(\langle\qvec{k},\qvec{R}_i\rangle) \times \mr{sin}(\langle\qvec{k},\qvec{R}_j \rangle)|,
	\end{split}
	\end{equation}
where $\bm{\hat{k}}$ is the unit vector of the direction in the sky under consideration, and $\qvec{R}_i$ and $\qvec{R}_j$ are the 
positions of the two satellites making up the baseline. To facilitate the data communication between the satellites and relative position measurement, the daughter satellites are tuned to have a stable configuration in the rotating frame, with the $z$-axis of each satellite always pointing to the Moon center, and one edge of the satellite oriented along the direction of the orbit.  Here in the simulation, as a simple model we will consider only one antenna, which we assume to be pointing towards the center of the Moon, so that its center of the beam lies on a plane perpendicular to the radial line from the lunar center to the satellite.

\subsubsection{The Moon Blockage}
\label{sec:shade}

\begin{figure}
	\centering
	\subfigure[ Without precession]{
    	\label{fig:shade_effect_nopre}
		\includegraphics[width=0.85\columnwidth]{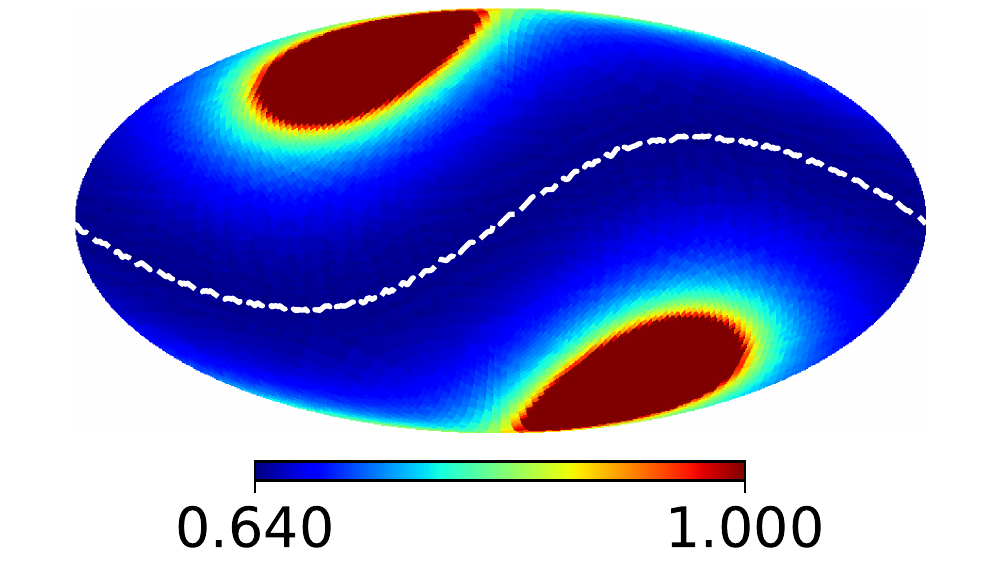}}
	\subfigure[ With precession ]{
    	\label{fig:shade_effect_pre}
		\includegraphics[width=0.85\columnwidth]{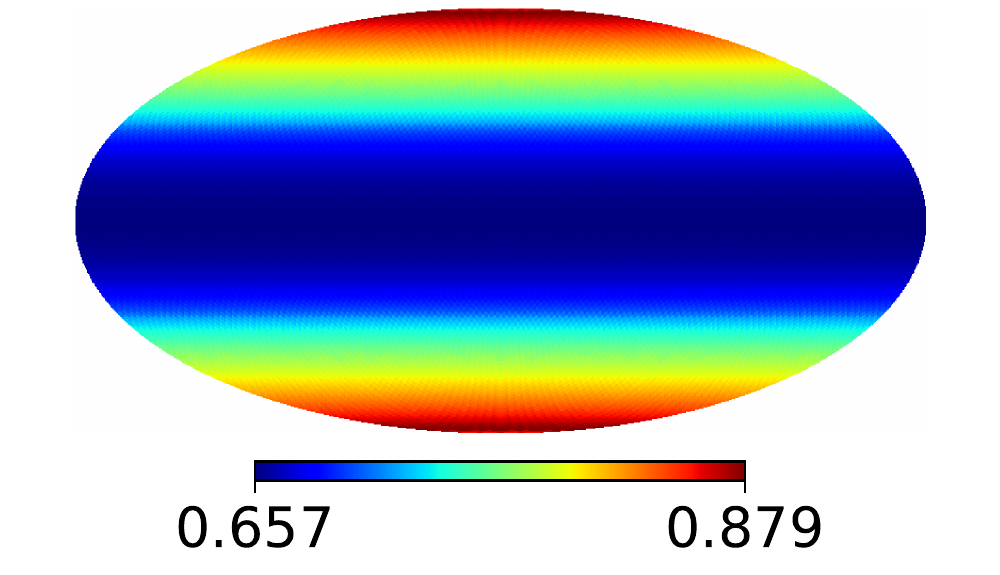}}	
	\caption{The visible probability of the sky for 5 satellites in the optimized distribution as shown in Fig.~\ref{fig:baseline_5s}.
	The {\it top panel} shows the case for no precession, and the white dotted line indicates the orbital plane of the 5 satellites.
	The {\it bottom panel} shows the visible probability with the orbit precession taken into account, and the nodal precession period is 1.3 years. 
	}
	\label{fig:shade_effect}	
\end{figure}

For a lunar orbit array, the Moon will block part of the sky at any time and the blocked regions vary with time as the satellites move in the orbit.
This is described by a shade function $S_{ij}(\bm{\hat{k}},t)$. 
For any given instant, the blockage function in Eq.(\ref{eq:Vij-2}) is determined by the two blocking functions w.r.t. the two satellites forming the specific baseline, i.e.
 \begin{equation}
	S_{ij}(\bm{\hat{k}}) = S_i(\bm{\hat{k}},\qvec{R}_i) \cdot  S_j(\bm{\hat{k}},\qvec{R}_j)
\end{equation}
where $S_i(\bm{\hat{k}},\qvec{R}_i)$ is the shade function for the $i$th satellite. $S_i(\bm{\hat{k}},\qvec{R}_i)$ is 1 for the unblocked sky and 0 for the blocked sky direction $\bm{\hat{k}}$.
In this work, we explicitly compute the shade function for each satellite at each time-step during the mock observation, and account for the fact that the blocked sky is slightly different for the different satellites in the array.

Fig.~\ref{fig:shade_effect} shows the probability of being visible to the satellites for different regions in the sky,
with the upper panel showing the case without precession, and the lower panel for the realistic case with nodal precession.

\begin{figure}
	\centering
	\includegraphics[width=0.35\textwidth]{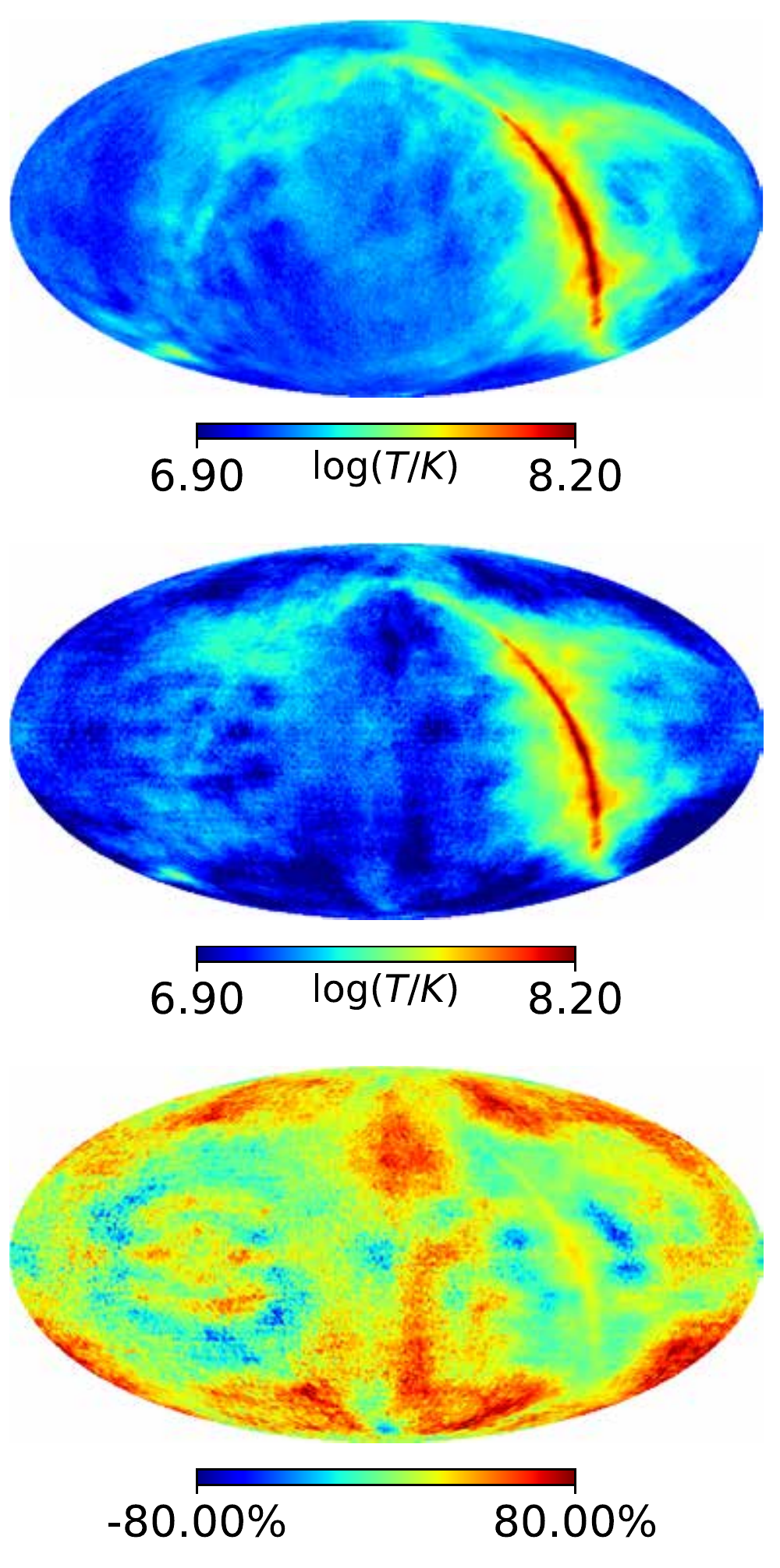}
 	\caption{The importance of taking the shading of Moon into account. Top Panel: A reconstructed sky map with the Moon blockage taken into account; Center Panel: A reconstructed sky map with the Moon blockage not taken into account. Bottom: Relative error between the two.}
	\label{fig:shade_effect_in_data_analysis} 
\end{figure}

In the reconstruction, it is vital to take the Moon blockage into account. We may test this by the following simple experiment: generate the mock visibilities with the Moon blockage, but then reconstruct the sky image without considering this, just as if the Moon is not present. The result is shown in Figure \ref{fig:shade_effect_in_data_analysis}. We see although the brightest part of the Galactic plane is still present in the reconstructed image, there are large errors in the map, especially at the low intensity regions. The method of inversion of  linear mapping offers however a general framework in which the blockage effect can be easily tackled with.  

In addition to the shielding effect, the Moon also reflects and emit radio waves. At the low frequency band we are considering, the thermal emission from the Moon itself is unimportant. We neglect the reflection effect in the present work and reserve it for future research.

\subsection{Computational Requirements}

\begin{table*}
\centering
\caption{Computational requirements of full-sky imaging for 1$\degree$ resolution ($N_{\rm side}=64$). }
\label{tab:computation}
\begin{threeparttable}
\begin{tabular}{cccccc}
\hline
 \makecell[c]{$\nu$ \\ $[\MHz]$}   & \makecell[c]{Integration \\ time [s]} & \makecell[c]{$\mathbf{B}$ size \\ $(\mr{N_{bl}}\cdot \mr{N_t}, \mr{N_{pix}})$}  & \makecell[c]{$\mathcal{B}_{l m}^1$ size in \\spherical harmonic space} & \makecell[c]{Cost \\ $[$core$\cdot$hour$]^2$}  \\
\hline 
1.5  & 26.26   & \makecell[c]{$(10\times 1.6\times10^6,49152)$\\(5.7 TB)} & \makecell[c]{$(10\times 1.6\times10^6,16471)$\\(1.9 TB)}
& $ 3.2\times10^4$\\
10   & 3.94 & \makecell[c]{$(10\times 1.04\times10^7,49152)$\\(38.1 TB)} & \makecell[c]{$(10\times 1.04\times10^7,16471)$\\(12.8 TB)}
&$ 2.1 \times10^5$\\
	\hline
	\end{tabular}
	\begin{tablenotes}
	\footnotesize
	\item [$^1$]The $\mathcal{B}_{l m}$ matrix 
	is a $(\mr{N_{bl}}\cdot \mr{N_t}, \mr{N_{lm}})$ matrix in the spherical harmonic space \citep{2018AJ....156...43H}. The angular resolution of the array requires $l_{\mr{max}} \geq \pi/\theta_0$, where $\theta_0$ is the angular resolution, and  $l_{\mr{max}}=180$ for $1^\circ$ resolution.
	\item [$^2$] The unit of $[$core$\cdot$hour$]$ represents a CPU hour on 1 core. Here we used 5.3 days with 250 CPU cores for each simulation at 1.5 MHz.
	\end{tablenotes}
	\end{threeparttable}
\end{table*}

The computational cost of the imaging simulation is dominated by the multiplication and inversion of the matrices.
For a single frequency, the beam matrix $\mathbf{B}$ in Eq. (\ref{eq:B}) has a dimension of $(\mr{N_{bl}}\cdot \mr{N_t}, \mr{N_{pix}})$. 
For long-term observations, $\rm N_{\rm bl}N_{\rm t}\gg \mr{N_{pix}}$, and it is beyond the capability of most current computers to compute the pseudo-inverse of matrix $\mathbf{B}$.
Therefore, we choose to reconstruct the sky map through Eq.(\ref{eq:T1}), by computing $\mathbf{(B^{\mr{H}}N^{-1} B)^{-1}}$ instead of $\mathbf{B^{-1}}$ as in \cite{2018AJ....156...43H} by the Moore-Penrose pseudo-inverse, then the cost of the matrix inversion scales as $\mr{N_{pix}}^3$.
However, the computation of the matrix $\mathbf{(B^{\mr{H}}N^{-1} B)}$ scales as $(\mr{N_{pix}}\mr{N_{bl}}\mr{N_t}+\mr{N_{pix}}^2\mr{N_{bl}}\mr{N_t})$ (where $\mr{N_t}N_{\rm bl}\gg \mr{N_{pix}}$), then the matrix multiplication dominates the computational cost in the entire algorithm. 

In order to make the computation feasible, we divide the entire observation period into multiple time steps in simulating the imaging. For example, imaging at 1.5 MHz can be performed every 15 days, then totally 30 time-steps in one precession cycle (1.3 years) are required. 

Even with such a step-by-step imaging approach,  we can produce the full-sky map only at a resolution of $1^\circ$ (corresponding to  $N_{\mr{side}}= 64$ in Healpix), which is cruder than the resolution of the array, e.g. 6.5 arcmins at 1.5 MHz for the maximum baseline of 100 km, because the computational cost for the inversion of the matrix $\mathbf{(B^{\mr{H}}N^{-1} B)}$ scales as $\mr {N_{pix}}^3$, which is quite demanding beyond this resolution. We list the full-sky imaging simulation requirement on computing resources in Table~\ref{tab:computation}. One will need more computing resources to achieve higher resolutions.  

In order to overcome the difficulty in achieving the resolution corresponding to the maximum baseline of the DSL array, we may apply the map-making algorithm to a small patch of the sky. This is detailed in the next subsection, and we can then estimate the point source sensitivity of the array.

\subsection{Part-sky Reconstruction}
\label{sec:partial}
\begin{figure}
	\centering
	\subfigure[Partial sky in full reconstructed map]{
    	\label{fig:mask_in_full}
		\includegraphics[width=0.9\columnwidth]{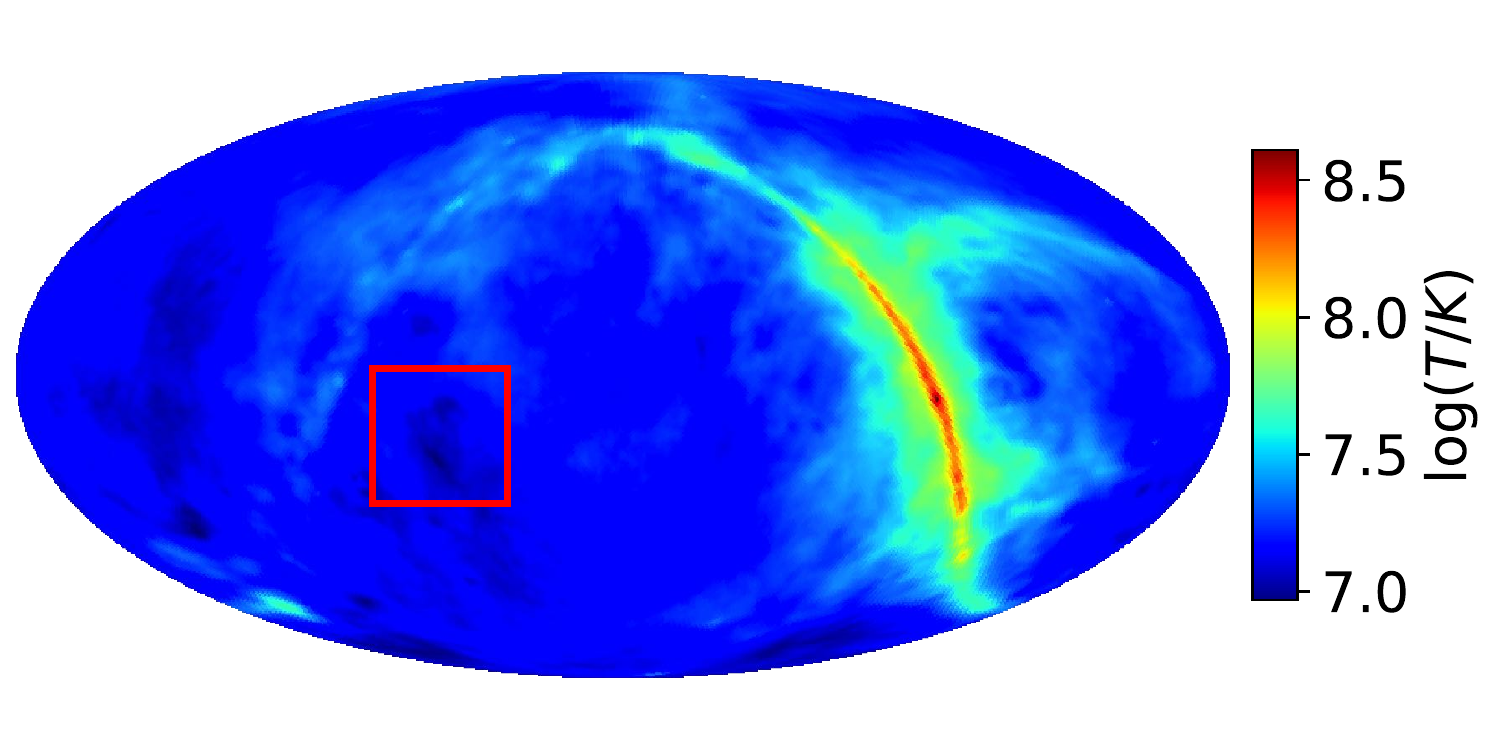}}
	
	\subfigure[Partial sky reconstructed map]{
    	\label{fig:mask_output}
		\includegraphics[width=0.9\columnwidth]{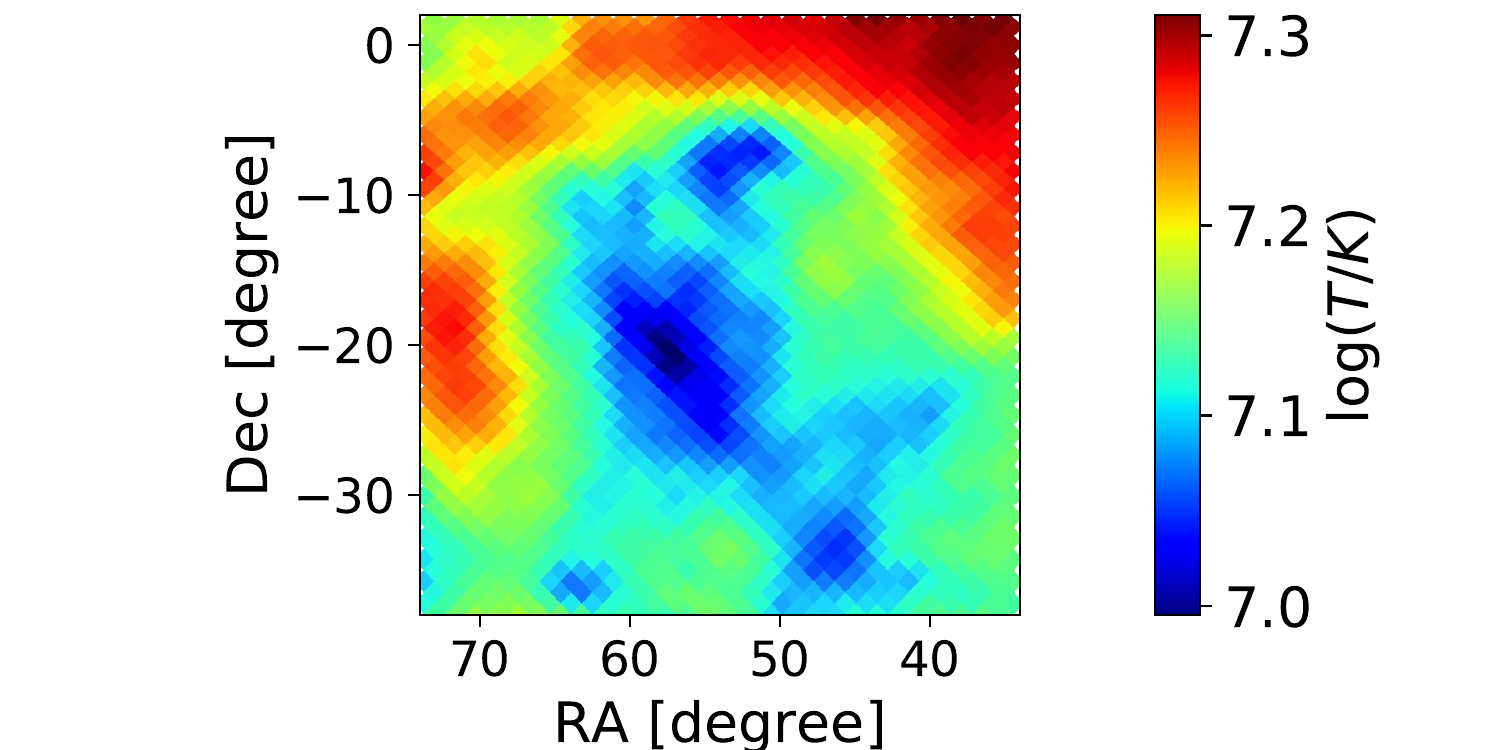}}		
	\subfigure[Relative errors between (a) and (b)]{
    	\label{fig:mask_full_relative_error}
		\includegraphics[width=0.9\columnwidth]{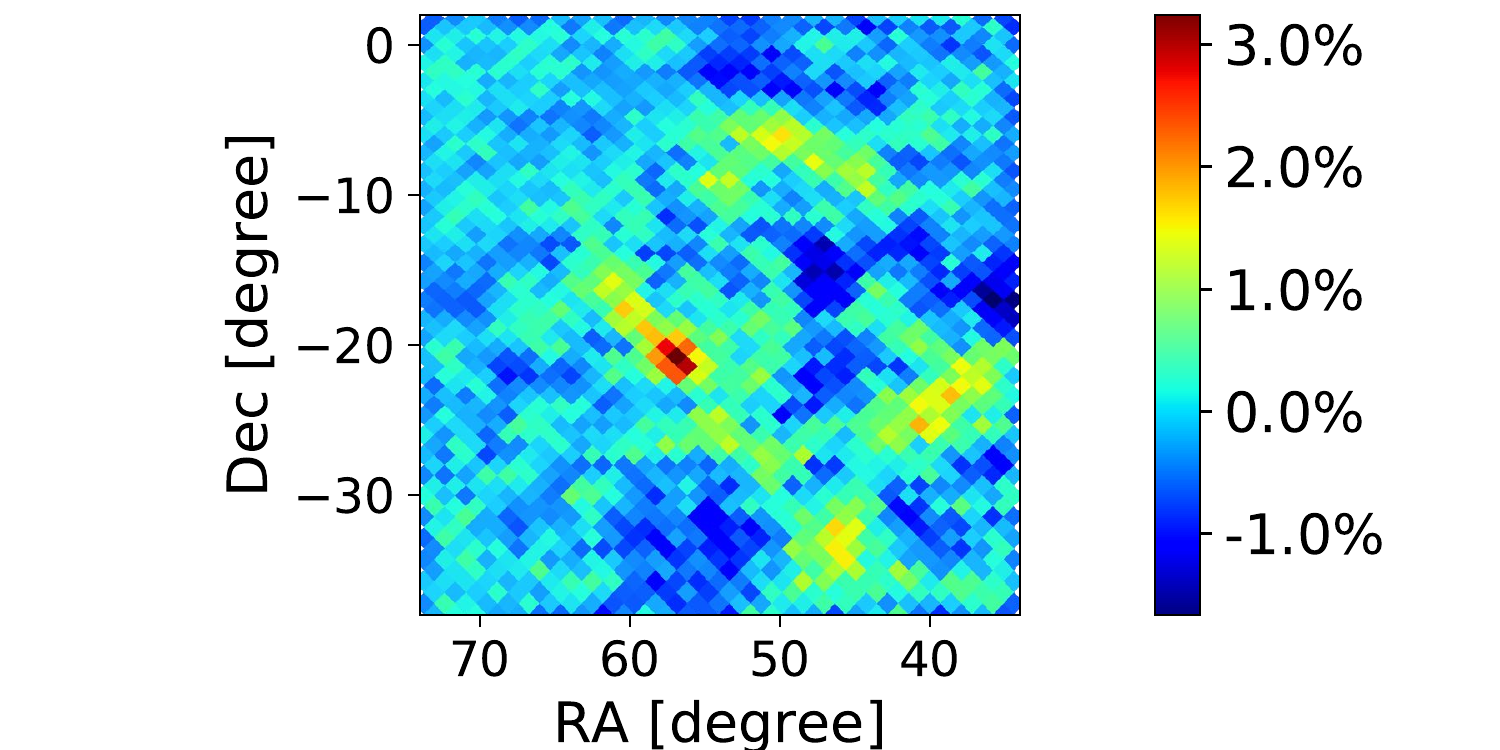}}	
	\caption{
	The {\it top panel} shows the reconstructed map from full-sky imaging without noise. Here we marked the part-sky region (RA:[34$^\circ$, 74$^\circ$], DEC:[-38$^\circ$, 2$^\circ$]) in red box.
	The {\it middle panel} shows the reconstructed map from part-sky reconstruction and under the same conditions as the full-sky imaging.
	The {\it bottom panel} shows the relative errors between full-sky reconstruction and  part-sky reconstruction.
	The total observation time is one precession period (1.3 years), the integration time is 26.26 s, and the $uvw$ coverage formed by the 5 satellites is shown in Fig.~\ref{fig:baseline_5s}.
	}
	\label{fig:compare_full_and_part}	
\end{figure}

For the maximum baseline of 100 km, the resolution is 10.31 arcmins at 1.0 MHz (corresponding to $N_{\rm side} = 512$ and $3\times 10^6$ pixels in the whole sky), and 0.344 arcmins at 30 MHz (corresponding to $N_{\rm side} = 16384$ and $3\times 10^9$ pixels in the whole sky). However, with the huge number of pixels in the full-sky map, the computation needed for reconstructing the full-sky map is quite formidable. The computation can be greatly reduced by restricting the simulation and reconstruction to a small patch of the sky. Now in the real observation, the visibility will also receive contribution from the sky outside the patch. In the reconstruction, this may produce some error in the reconstructed map. We shall call this the ``sidelobe effect". It is not due to the side lobe of the antenna, whose main lobe is much wider than the region of concern, rather it simply refers to the fact that in the part-sky reconstruction the radiation outside the patch may also affect the reconstruction result. 
We assess this side-lobe effect by checking the consistency between the imaging results for a small region of the sky, with the corresponding sky region from the full-sky imaging.

For a small patch of the sky,  we set the pixels outside this patch of sky to be 0, i.e.
\begin{eqnarray}
\mathbf{V_p} = \mathbf{B_p T_p + n_p}.
\label{eq:matrixVp}
\end{eqnarray}
For a total of $N_{\mr{t}}$ observation time points, $\mathbf{V_p}$ is still a vector of $(N_{\mr{t}} \cdot N_{\mr{bl}})$, but the computational cost of $\mathbf{B_p}$, $\mathbf{(B_p^\mr{H} N_p^{-1} B_p)}$, and its inverse is largely reduced, as $N_{\mr{pix}}$ is now given by the pixel number of this patch of sky. The expected value of the estimator is:
\begin{eqnarray}
\mathbf{\langle{\mathbf{\hat{T}_p}}\rangle} = \mathbf{(B_p^\mr{H} N_p^{-1} B_p)^{-1} B_p^\mr{H} N_p^{-1} B_p T_p}.
\end{eqnarray}

Fig.~\ref{fig:compare_full_and_part} compares the results of the full-sky reconstruction from the entire orbit precession period (1.3 year) at $1^\degree$ resolution (panel (a)) with the part-sky reconstruction for the region in the red rectangle (right ascension (RA): [34$^\circ$, 74$^\circ$], declination (DEC): [-38$^\circ$, 2$^\circ$]) under the same observation conditions (panel (b)). In order to eliminate the sidelobe effect, we conservatively use the central quarter of the simulated sky area to estimate the array sensitivity in the part-sky simulation.
The relative errors are plotted in panel (c). 
The results obtained by the two approaches are quite similar, with a relative difference less than 5\%. This shows that the sidelobe effect is not significant for this area of the sky, and the part-sky reconstruction approach is feasible.

\begin{figure*}
    \subfigure[10 MHz: ideal situation.]{
    \includegraphics[width=0.6\columnwidth]{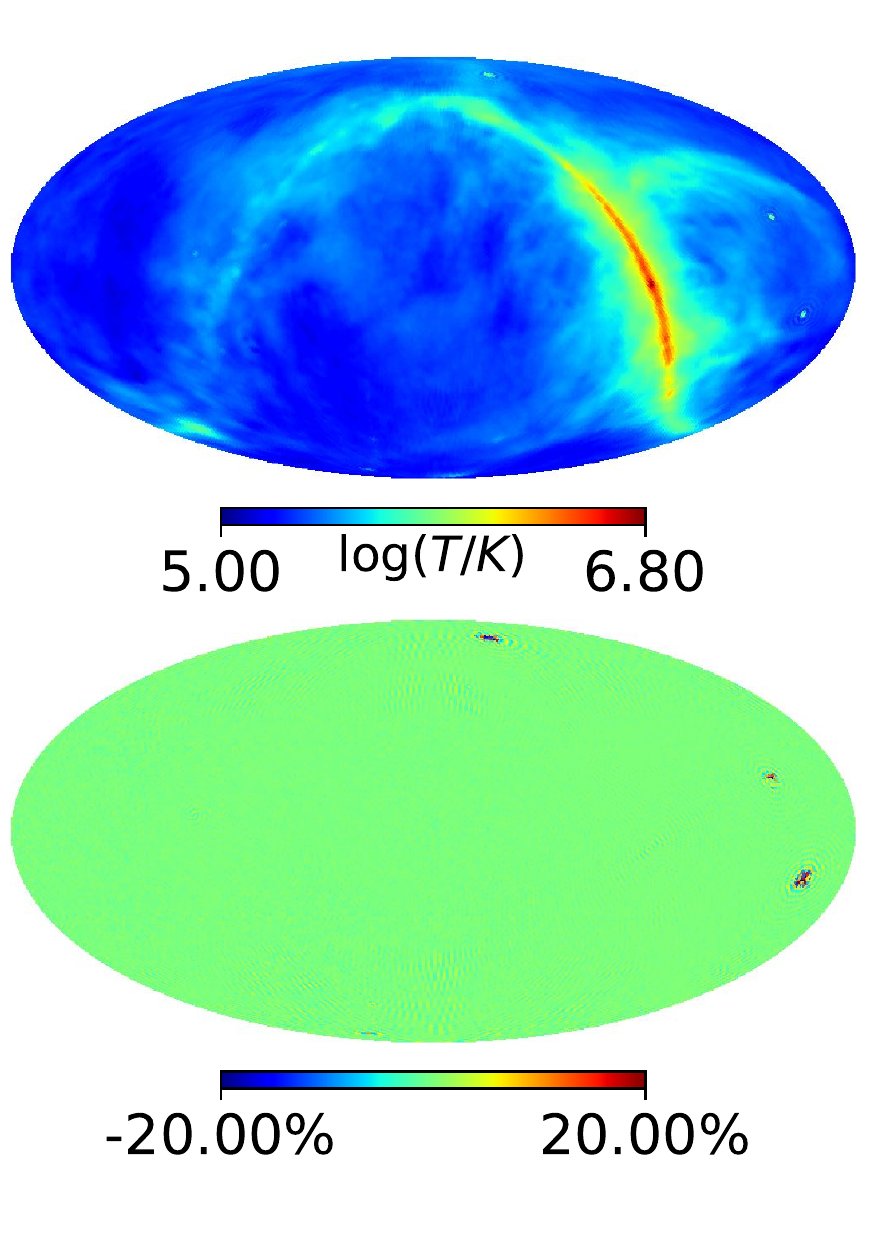}}
    \quad \quad 
    \subfigure[10 MHz: with noise.]{
    \includegraphics[width=0.6\columnwidth]{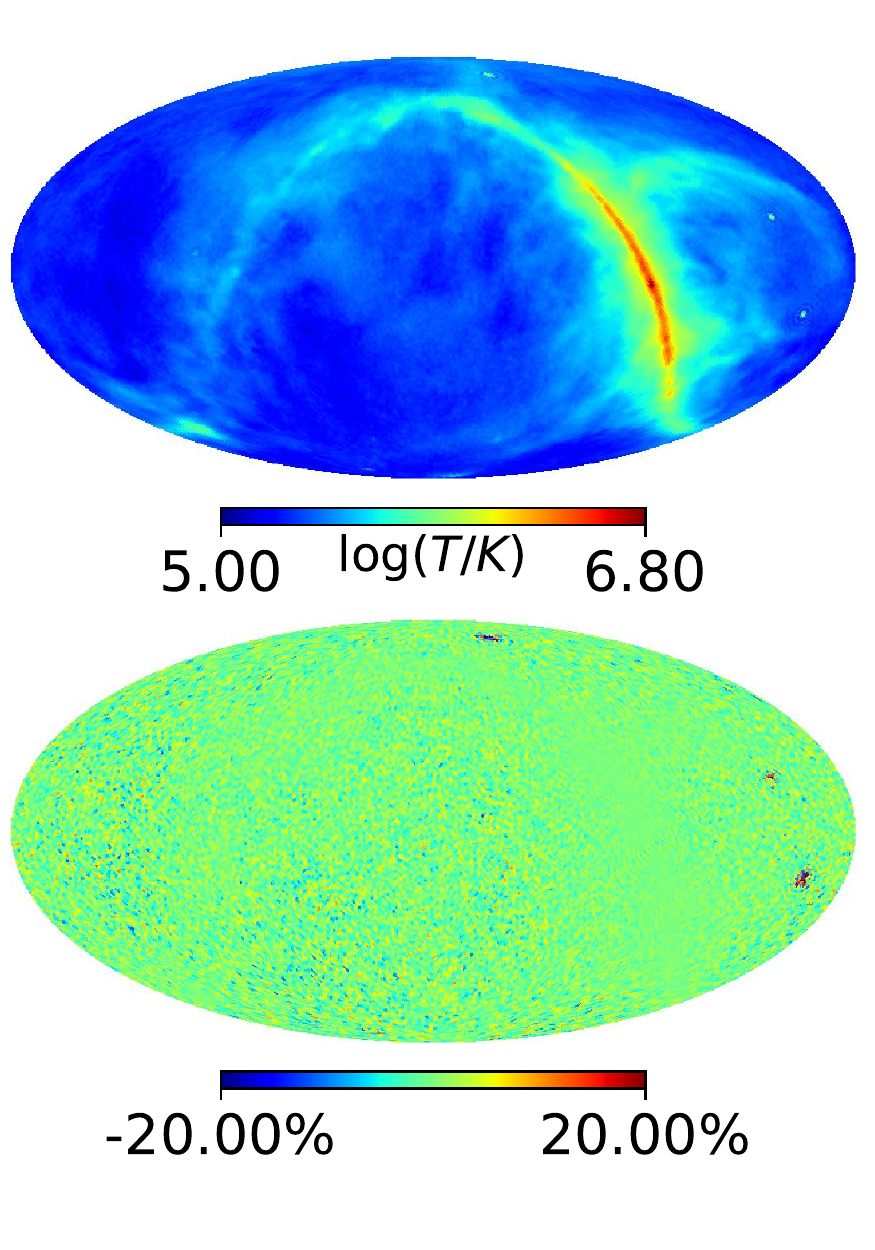}}   
    \quad \quad
    \subfigure[10 MHz: with Moon blockage and noise.]{
    \includegraphics[width=0.6\columnwidth]{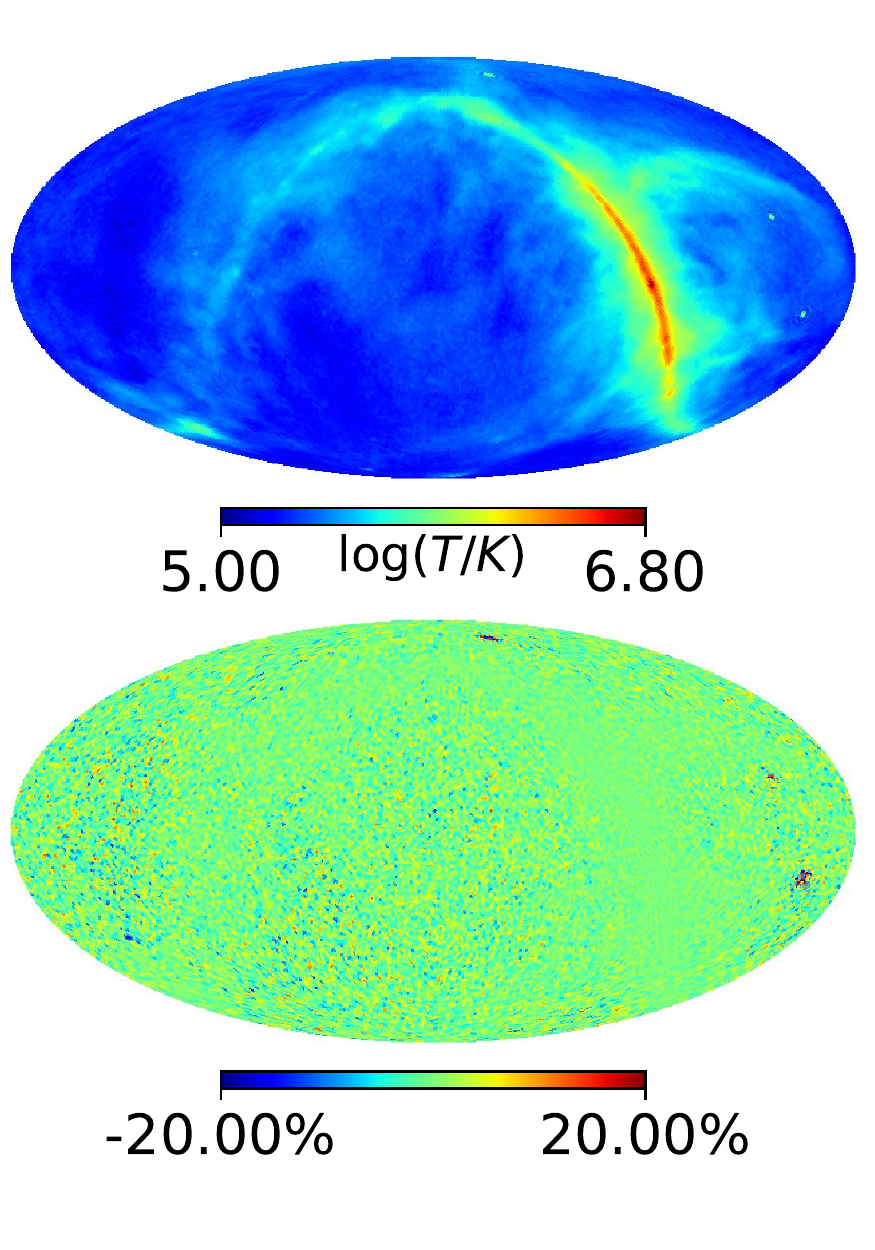}}   
    \quad \quad
    \caption{The {\it upper panels} show the reconstructed sky maps at 10.0 MHz for the cases of (a) $\sim$ (c), and the {\it lower panels} show the relative error maps. 
	}
	\label{fig:full_sky_results10MHz}
\end{figure*}

\section{Simulation Results}
\label{sec:results}

We can now generate the mock visibilities from the input sky map with realistic beam and noise as Eq.(\ref{eq:matrixV}), and then reconstruct the sky map using the estimator given by Eq.(\ref{eq:T1}). Then we evaluate the quality of the reconstructed map and the sensitivity of the array that can be reached after a reasonable time of observation. Below, we shall first present our simulation  results obtained with different observational conditions, then assess the overall quality of the image by a few different statistical measures, and finally we estimate the sensitivity for point sources. As representative examples, we study the results of image synthesis at $\nu=1.5 \MHz$ and 10 MHz, respectively. With the maximum baseline of 100 km, the angular resolution could reach $\sim 7'$ at 1.5 MHz, and $\sim 1'$ at 10 MHz. As discussed earlier, making a simulation with such a resolution for the whole sky requires a large amount of computation. Below, we shall first study the overall imaging quality of a full-sky map with a resolution of $1^\circ$, and then investigate the point spread function with a higher angular resolution using the part sky reconstruction result. 

To minimize the data rate to be processed on orbit, the array could be designed to have different integration times for baselines with different lengths. For example, with the array of 5 elements in the ``optimized spacing'' configuration, the 10 baselines have lengths of 1.0 km, 9.0 km, 10.0 km, 21.6 km, 30.6 km, 31.6 km, 68.4 km, 90.0 km, 99.0 km, and 100.0 km, respectively, and the maximally allowed integration time at 1.5 MHz are then 26.26 s,  2.92 s,  2.63 s,  1.21 s,  0.86 s,  0.83 s,  0.38 s,  0.29 s,  0.27 s, and 0.26 s, respectively, according to the upper limit of Eq.(\ref{eq:int_time}). However, to reduce the computation load, in the following we use a uniform integration time, chosen according to the 1 km baseline, for all baselines, i.e. we set the integration time to be 26.26 s at 1.5 MHz and 3.94 s at 10.0 MHz. Then the $uvw$ space is under-sampled as compared to the real case. We have scaled down the thermal noise on each visibility measurement accordingly, but nevertheless this simplification will still have some adverse effects on the image synthesis, because fewer $uvw$ points are sampled, degrading the imaging quality.
This is especially significant for short observation time, when the amount of precession of the orbital plane is still small, and the under-sampling in the $uvw$ space has a more apparent effect.
We have tested the result of the reconstructed sky in 10 days' observation using this simplified integration time, comparing with the one using the actual setting of the integration time, and found that 
the mean square error (defined below by Eq.(\ref{eq:MSE})) is increased by 32\%.
As time passes, the effect of under-sampling decreases. By using the uniform integration time, we will get more conservative estimates in both the global imaging quality and the point source sensitivity.

\subsection{Global Map}
\label{sec:results_global}

We shall start with a simple, idealized case, and then add practical issues one by one to see how each factor affects the result. 

{\bf (a)} {\it The Ideal Case.}  This is the case that is free of noise and Moon blockage, and the antennas are assumed to have isotropic response.
These conditions are similar to those adopted in  \citet{2018AJ....156...43H}, though there the visibility data were randomly sampled in the orbit-allowed region of the $uvw$ space, whereas here they are generated from the orbital motion with a given array configuration.   

{\bf (b)} {\it Noise included.} Based on the ideal case, we first add a Gaussian random noise, with the standard deviation of $\sigma_{\rm n}$ given by Eq.~(\ref{eq:noise}), to the mock visibility data, so that the sensitivity will be limited by the noise level. 

{\bf (c)} {\it Effect of Moon blockage.} Inevitable for a lunar orbit array, the Moon blockage makes the visibility dependent on the position of the satellites on the orbit. We have noted earlier that this effect has to be taken into account in the reconstruction. Furthermore, it also reduces the effective observation time. Here we take into account the time-varying Moon blockage.

\begin{figure}
\centering
\includegraphics[width=0.8\columnwidth]{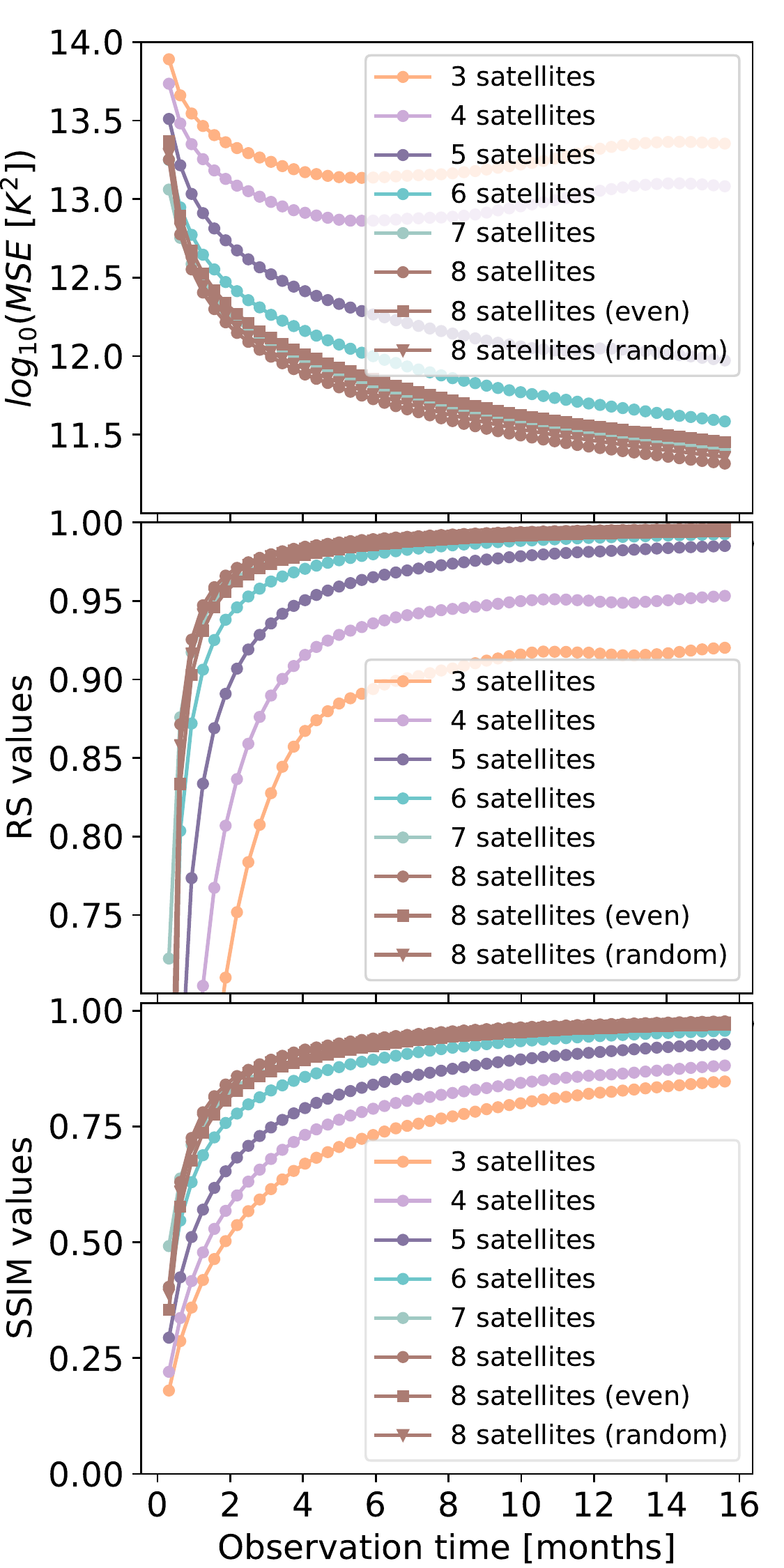}
\caption{The MSE ({\it top panel}), RS value ({\it middle panel}), and SSIM ({\it bottom panel}) of 
reconstructed sky maps at 1.5 MHz as a function of observation time.  The different lines with dots in different 
colors correspond to different numbers of satellites in optimized spacing configuration as indicated in the legend.
In the case of 8 satellites, we also show the results for even spacing (brown lines with squares) and random spacing (brown lines with triangles) configurations for comparison.}
\label{fig:full_sky_results2}
\end{figure}

\begin{figure}
	\centering
	\includegraphics[width=0.7\columnwidth]{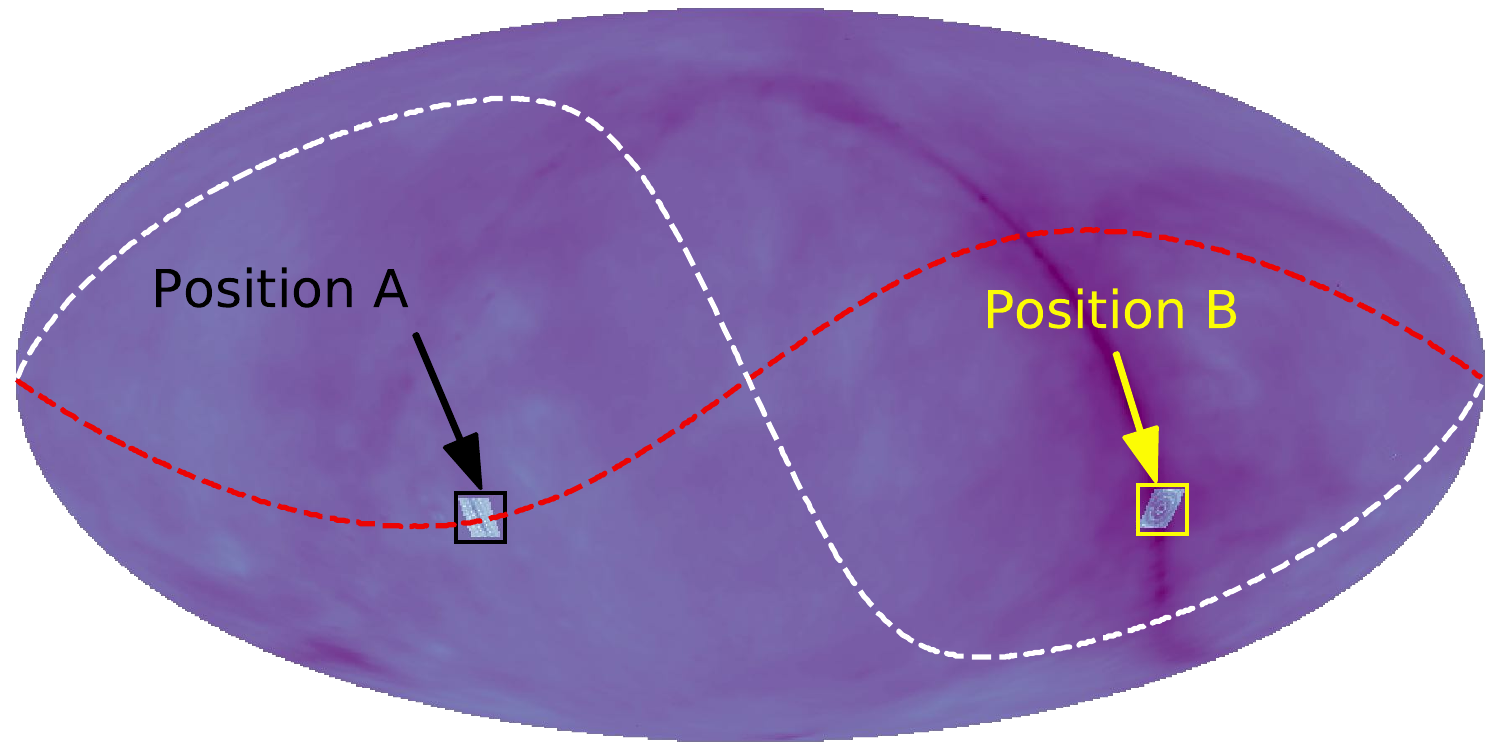}
	\caption{The positions of two patches in the sky. The red dashed line shows the orbital plane with $30^{\circ}$ inclination and the white dashed line indicates the plane perpendicular to the orbital plane at the initial time. Here we have selected two special positions, \textit{position A} (RA: $[68^\circ, 76^\circ]$, DEC: $[-32.8^\circ, -24.8^\circ$]) in black rectangle on the orbit plane, and \textit{position B} (RA: $[248^\circ, 256^\circ]$, Dec: $[-31.0^\circ, -23.0^\circ$]) in yellow rectangle away from the orbital plane. 	}
	\label{fig:psf_position}
\end{figure}

The reconstructed sky maps for these cases are shown in the upper panels of  Fig.~\ref{fig:full_sky_results10MHz} for the 10 MHz simulations.
We also show in the bottom panels in each figure the corresponding relative error maps, defined as 
$\varepsilon=(T-\hat{T})/T,$
where $T$ and $\hat{T}$ are the brightness temperatures of a pixel in the input sky map and the corresponding pixel in the reconstructed map, respectively.

\begin{figure*}
    \centering
    \includegraphics[width=1.5\columnwidth]{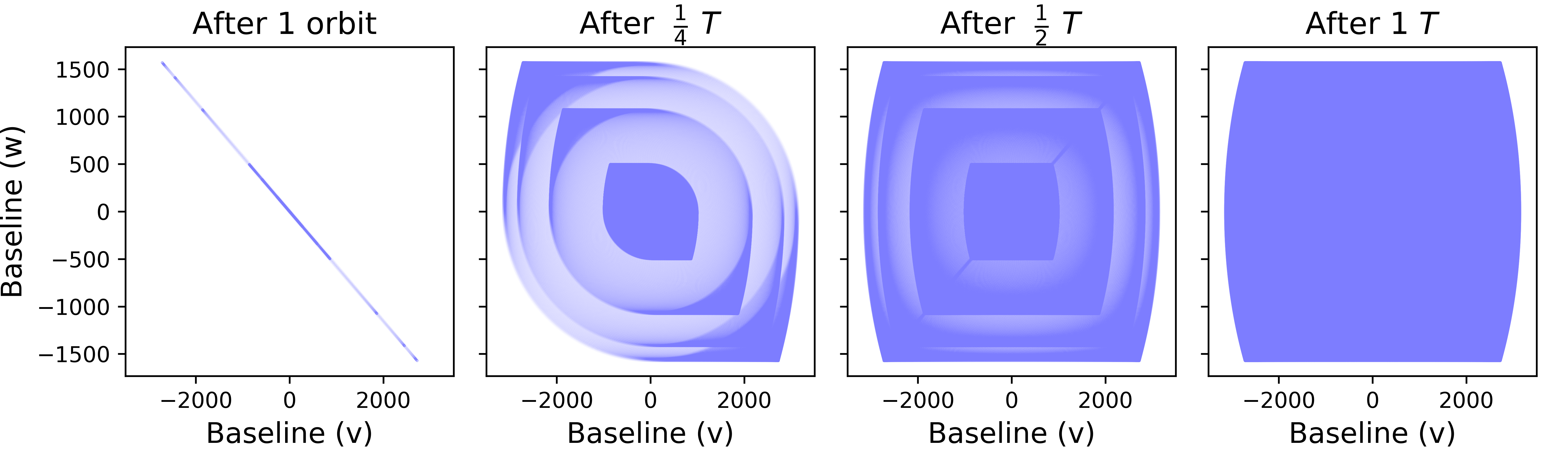}
    \includegraphics[width=1.5\columnwidth]{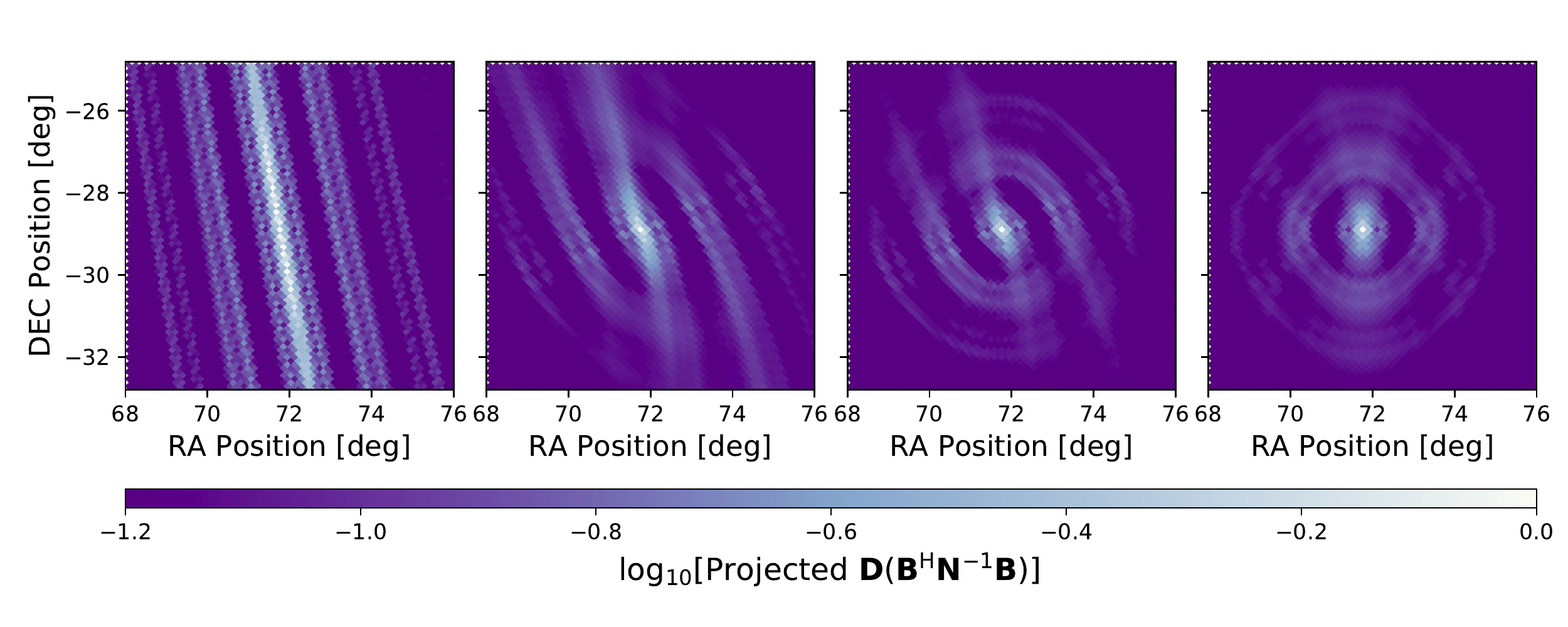}
    \caption{The projected $vw$  baseline coverage (top row) and the corresponding PSF (bottom row) for a lunar orbit linear array of 5 satellites in optimized configuration. From left to right,  the satellites complete the observation for 1 orbit ($8.25\times 10^3 $seconds), $1/4 \ T$, $1/2 \ T$, and $1 \ T$, where $T$ is the precession period of the orbital ascending node and equals to 1.3 years for the DSL mission.
 The sky region is chosen at the orbit plane at position A as illustrated in Fig. \ref{fig:psf_position}.
	}
	\label{fig:psf_orbit}
\end{figure*}

In all cases, the visual impression is that the sky is reasonably reconstructed, the error is only prominent at some pixels.  As expected, in the ideal case (a) the reconstructed map has very low errors. The presence of noise (case (b)) visibly increases the relative error in the map. The error increases further with the introduction of the Moon blockage (case (c)), which is expected because at least it reduces the observation time for each pixel.

To evaluate the global quality of the maps more quantitatively, we introduce a few commonly used statistics, such as the Mean Squared Error (MSE), the R Squared value (RS), and the Structural SIMilarity (SSIM). The MSE is a representation of the absolute error, while the RS value and the SSIM are normalized and quantify relative errors. These are widely used for image comparison. The MSE is defined as
\begin{equation}
{\rm MSE}(I,R) = \frac{1}{n_{\mr{pix}}} \sum_{i=0}^{n_{\mr{pix}}} [I(i)-R(i)]^{2},
 \label{eq:MSE}
\end{equation}
where $I$ and $R$ denotes the values of the input and reconstructed maps respectively. The MSE is the variance in the case of an unbiased estimator.  
Note that the value of MSE depends on the absolute magnitude of the measured quantity.
The RS value is related to the MSE by
\begin{equation}
 {\rm RS}(I,R) = 1-\frac{{\rm MSE}(I,R)}{\mr{Var} (I)}.
 \label{eq:RS}
\end{equation}
Its value ranges from 0 to 1, and is independent of the absolute magnitude of the measured quantity
The structural similarity is defined as
\begin{equation}
{\rm SSIM}(I,R)=\frac{(2 \bar I \bar R+c_{1})(2\, \rm{Cov}(I,R)+c_{2})}{(\bar I^2+\bar R^2+c_{1})(\sigma(I)+\sigma(R)+c_{2})},
\label{eq:SSIM}
\end{equation}
where $ \bar{I}$ and $\bar{R}$ are the mean values of the input and the reconstructed maps, $\sigma(I)$ and $\sigma(R)$ are the standard deviations of $I$ and $R$ respectively, and $\rm{Cov}(I,R)$ is the cross-covariance between the input and the output. $c_1=(k_1 L)^2$ and $c_2=(k_1 L)^2$ are very small constants used to maintain stability of the computation \citep{2004Geode.120...75H}, in which L is the range of the sky map data. In our analysis, we adopt $k_1=0.01$,$k_2=0.02$.

The MSE, RS and SSIM values for the cases described above are listed
in Table \ref{tab:full_sky_results}.  As we depart from the ideal case (a), with more practical issues taken into account, the MSE value increases while RS and SSIM values decrease,  indicating that the imaging quality degrades with the inclusion of more practical issues.

In Fig.\ref{fig:full_sky_results2} we show the MSE, RS and SSIM values of the reconstructed maps at 1.5 MHz as functions of total observation time, for different numbers of satellites, and different array configurations. Here, we also consider observation time much shorter than the full precession period. Although the precession is incomplete, as shown by the curves, within the span of a few months it is already possible to produce a good whole-sky map. The error level quantified by the MSE drops with the time as $1/\sqrt{t_{\rm obs}}$ as expected, and the quality of the map would be improved continuously with the accumulation of more data with the passing of time.
Comparison of the different curves also shows that while the sky map could be produced even with 3 satellites, the noise as quantified by MSE will be reduced as the inverse of the number of satellites, and the image quality as quantified by the RS and SSIM will also be improved with the addition of more satellites. 

\begin{table}
\centering
\caption{Comparison of the full-sky imaging results with different approximations, for 1$\degree$ resolution map ($N_{\rm side}=64)$.}
\label{tab:full_sky_results}
\begin{threeparttable}
\begin{tabular}{cccccc}
\hline
 \makecell[c]{Case} & \makecell[c]{$\nu$\\ {[MHz]}} & \makecell[c]{Description} & \makecell[c]{MSE\\{[$\rm K^2$]}}  & \makecell[c]{RS} & \makecell[c]{SSIM} \\
\hline 
(a) & 1.5 & Ideal case    & $1.59\times10^{10}$  &1.000 & 1.000 \\
(b) & 1.5 & \makecell[c]{Noise} &  $2.59\times10^{11}$ &0.995 & 0.969 \\
(c) & 1.5 &  \makecell[c]{Noise, Moon blockage} & $3.59\times10^{11}$ &0.993 & 0.959\\
(a) & 10.0 &  \makecell[c]{Ideal case} & $1.03\times10^{7}$ & 1.000 & 0.997\\
(b) & 10.0 & \makecell[c]{Noise}&$4.55\times10^{7}$&0.996&0.972 \\
(c) & 10.0 & \makecell[c]{Noise, Moon blockage}& $5.74\times10^{7}$ & 0.995 & 0.964\\
\hline
\end{tabular}
\begin{tablenotes}
\footnotesize
\item Parameters in these simulations:\\ 
-- The integration time $t_{\mr{int}} =$ 26.26 s at 1.5 MHz, and $t_{\mr{int}} =$ 3.94 s at 10.0 MHz.\\
-- The total observation time $t_{\mr{obs}}=$ 1.3 year.\\
-- The system noise temperature is $5.4\times10^4$ K at 1.5 MHz with the corresponding integration time, and $2.0\times10^3$ K for 10.0 MHz.
	\end{tablenotes}
	\end{threeparttable}
\end{table}

\begin{figure}
\centering{
    \includegraphics[width=0.95\columnwidth]{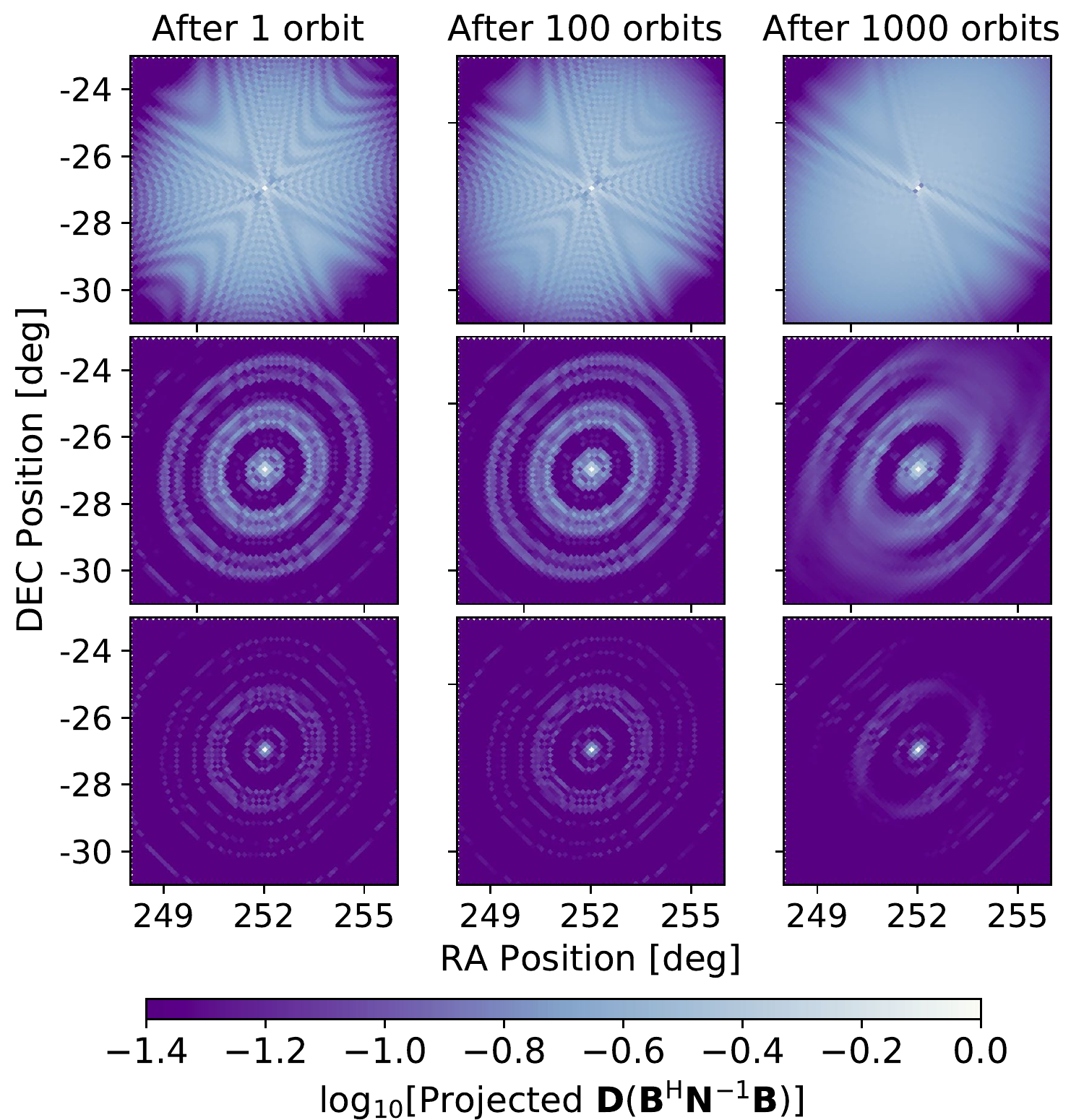}
    \caption{The PSF at position B for different numbers of satellites and different observation time. The three rows correspond to the PSF for 3, 5, and 7 satellites, from top to bottom respectively, and the three columns show from left to right the PSF maps for exposure time of satellites completing 1, 100, and 1000 orbits flight (8248.7 seconds for each orbit). }
    \label{fig:psf_sate}
    }
\end{figure}

We also compare the even spacing configuration, random spacing configuration, and the optimized configuration for the 8 satellites case in Fig.\ref{fig:full_sky_results2}. The result of 8 satellites with the optimized configuration is obviously better than the other two configurations and that of uniform distribution of satellites is the worst.

\subsection{PSF and eigenvalue comparison}
\label{sec:psf}

Due to the limitation of computer memory, in our study of the global image quality we have adopted a relatively low resolution ($1^\circ$). 
We now turn our attention to imaging a small part of the sky and analyze the resolution and imaging capabilities of the array, using the PSF and eigenvalue spectra \citep{2016ApJ...826..181D}. Our imaging algorithm can also deal with a part of the sky.  As an example, we selected two particular patches of sky with a higher angular resolution of 6.7 arcmin (HEALPix $N_{\mr{side}}=512$), shown in Fig. \ref{fig:psf_position}, i.e. \textit{position A} at RA $[68^\circ, 76^\circ]$, DEC $[-32.8^\circ, -24.8^\circ$], and \textit{position B} at RA $[248^\circ, 256^\circ]$, DEC $[-31.0^\circ, -23.0^\circ$]. 
Position A is located on the satellite orbit plane, while position B is located away from the orbital plane. Here we will show the PSF at position A to illustrate the influence of orbital precession on map-making, and the PSF at position B to illustrate the optimization of the satellite array configuration for map-making.

In this work, we plot the PSF for the dirty beam, $\mathbf{D(B^\mr{H} N^{-1}B)}$, where $\mathbf{D}$ is the normalization matrix.
In Fig.~\ref{fig:psf_orbit}, we show the results of PSF at position A after different exposure times (i.e. 1 orbit (8248.7 seconds), $1/4 \ T$, $1/2 \ T$, and $1 \ T$, from left to right respectively, where $T$ is the precession period) and the projected $vw$ coverage of the corresponding baseline distribution.  When the satellite completes a single orbit, the baseline is almost on a plane, and the PSF in the direction on the orbital plane is the linear stripes shown in the first lower panel of Fig.~\ref{fig:psf_orbit}. When the orbital precession completes 1/4 period, the linear striped side lobes are obviously weakened, and an incomplete ring-shaped structure appears at the same time. After 1/2 period observation, the ring-shaped stripe tend to be closed and the linear stripe structure is further weakened. After completing the whole cycle of precession, the PSF evolves to what is shown in the last panel in Fig.~\ref{fig:psf_orbit}. The side lobes at the border almost disappeared, and an X-shaped structure appeared at the center. This arise from the triangular gap at the upper and lower ends of the $vw$ plane shown in the upper right panel of Fig.~\ref{fig:psf_orbit}.

We compute the PSF for the $8^\circ \times 8^\circ$ facet at position B for 1, 100, and 1000 orbits' observation with different numbers of satellites, which is shown in Fig.~\ref{fig:psf_sate}. 
For simplicity, here we do not consider the shielding of Sun and Earth, but simply take all orbital time as the effective observation time. The results of 3 satellites (3 baselines), 5 satellites (10 baselines), and 7 satellites (21 baselines) are shown from top to bottom rows in Fig.~\ref{fig:psf_sate}.
Note that the PSF matrix contains all the relevant information about the telescope's ability to imaging the sky. It is clearly seen that the side lobes are more  suppressed with increasing number of satellites. Some of the ripples (or stripes) in the PSF is caused by the under-sampling of baselines in the torus envelope of the precessing orbital plane, and such ripples would be suppressed with longer observation time, as clearly shown in the first row of Fig.~\ref{fig:psf_sate}, but there are also stripes originated from the inadequate sampling in the upper and lower cone regions of the $uvw$ space(see the upper panels in Fig.~\ref{fig:psf_orbit}), as the $vw$ (or $uw$) cross-section is blank at the upper and lower ends. This structure will not disappear because it is caused by the peculiarity of the baseline configuration generated by the orbital inclination of 30$^\circ$ and the orbital precession.

\begin{figure}
\centering{
 \includegraphics[width=0.74\columnwidth]{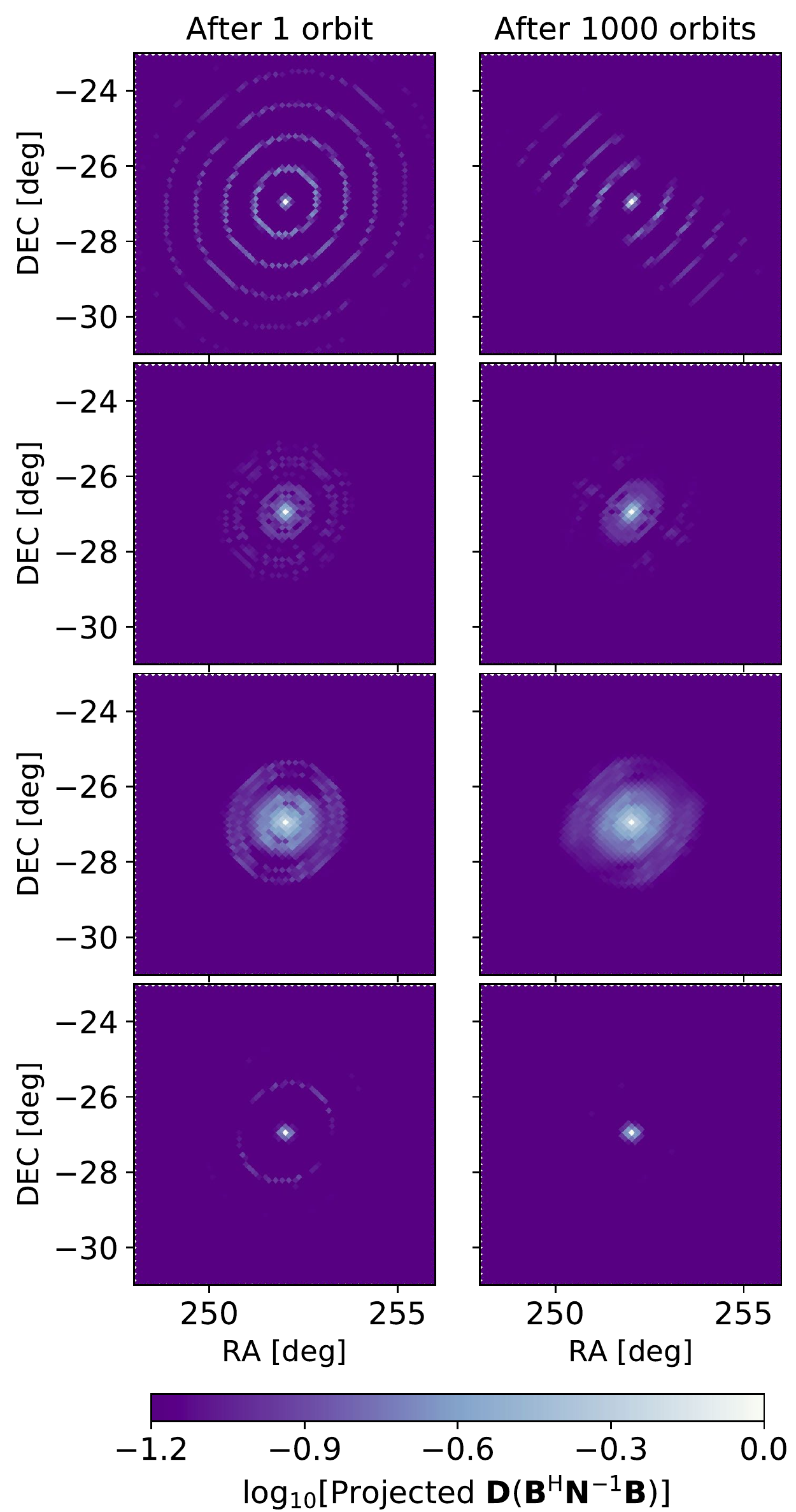}
\caption{The PSF for different arrangements of 8 satellites  and different observation time. The four rows from top to bottom correspond to even spacing (a), random spacing (b),   logarithmic spacing (c), and optimized spacing (d) configurations, respectively. The left and right columns are for the PSF after the exposure time corresponding to 1 and 1000 orbits (8248.7 seconds for each orbit), respectively.}
    \label{fig:psf_8sate}
}
\end{figure}

In Fig.~\ref{fig:psf_8sate}, we plot the PSF for different array configurations. From top to bottom, the four rows show the PSF formed by 8 satellites with even spacing, random spacing, logarithmic spacing, and optimized spacing configurations, respectively. The left and right columns show the PSF after 1 orbit and 1000 orbits observation respectively. 
It is seen that in the even spacing configuration,
the structure of the PSF is composed of multiple concentric rings, and the side lobes are more prominent, showing the effect of having more redundant baselines. The PSF of logarithmic configuration also has prominent side lobes around the center, because it has more short baselines as shown in Fig.~\ref{fig:bl_length}. Among the four configurations, the optimized spacing has the sharpest PSF, thanks to the largest number of long baselines. We also find that the side lobes in the PSF is reduced  considerably by prolonging the observation time. 

We also analyze the eigenvalue spectrum of the matrix $\mathbf{B^{\mr{H}}N^{-1}B}$, which is approximately the inverse of noise covariance matrix of the maximum-likelihood estimate. The matrix $\mathbf{B^{H}N^{-1}B}$ contains all information about the instrument correlation in our map, but we have an implicit assumption here that the noise is uncorrelated between pixels, so the matrix $N^{-1}$ is diagonal in our simulation. We anticipate that the smallest covariance will be associated with the greatest inverse covariance. 

\begin{figure}
\centering
\subfigure[Different number of satellites]{
\includegraphics[width=0.8\columnwidth]{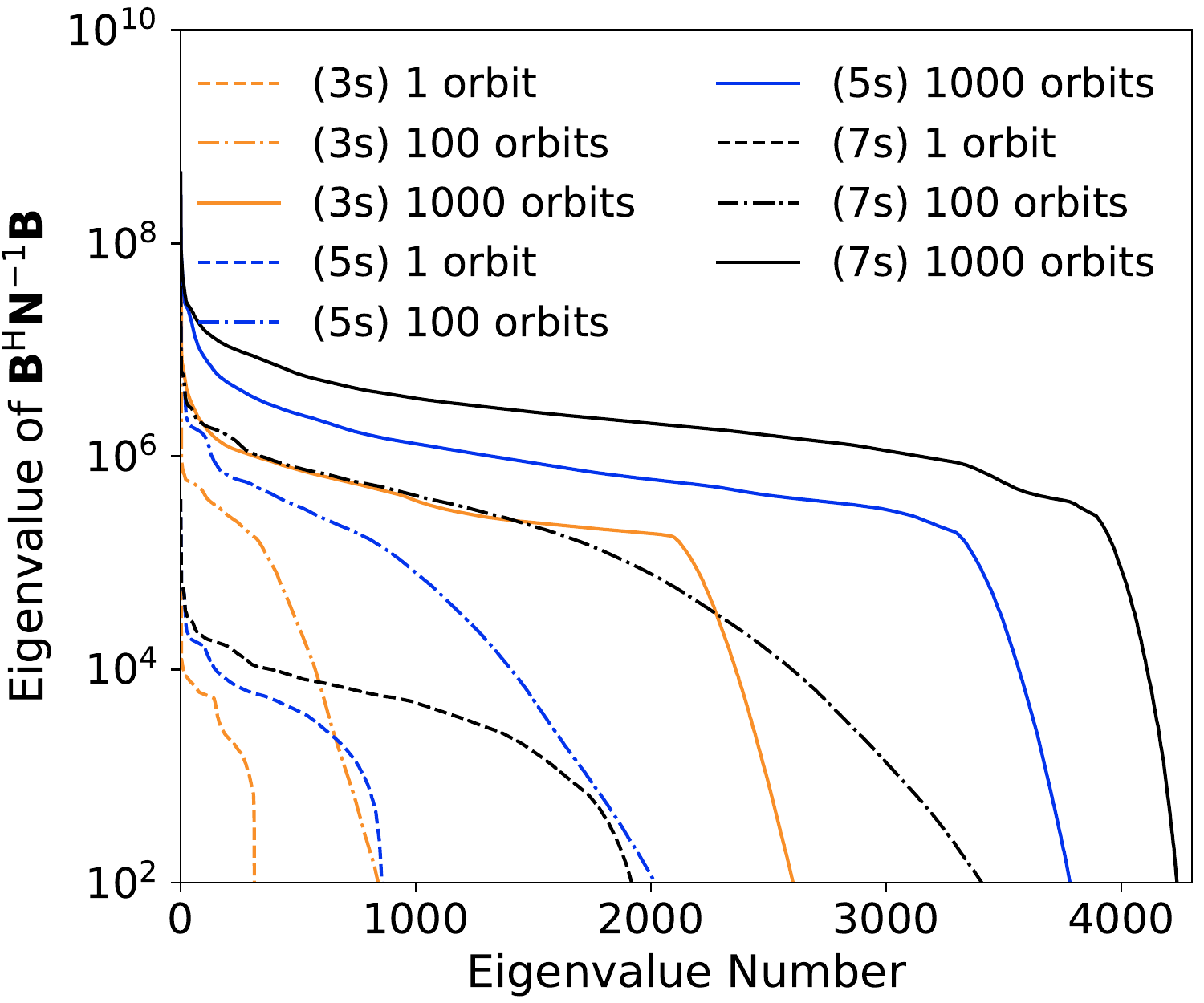}}
\subfigure[Different configuration of satellites]{
 \includegraphics[width=0.8\columnwidth]{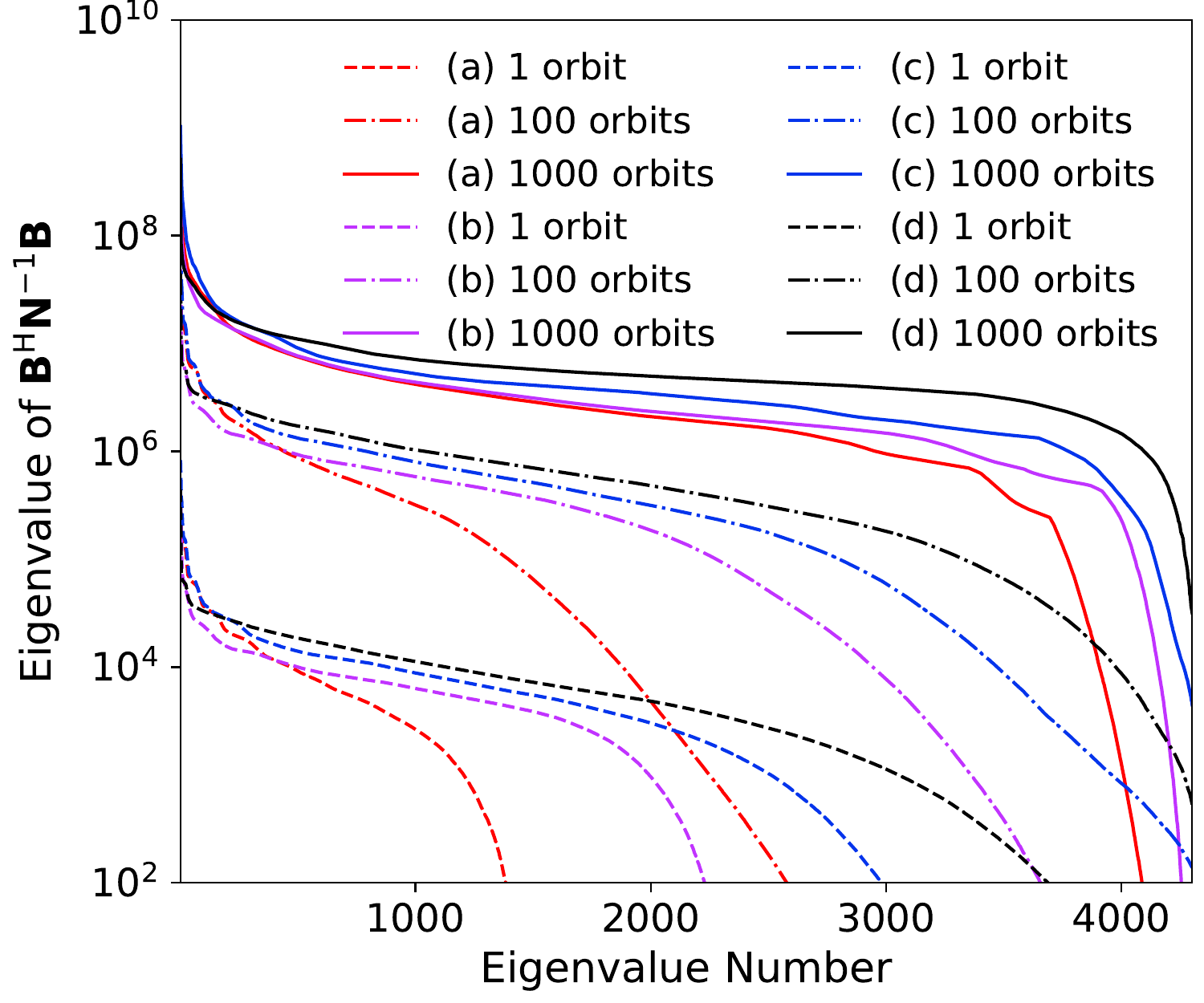}
}
    \caption{The eigenvalue spectra of $\mathbf{B^{\mr{H}}N^{-1}B}$. The {\it upper panel} shows the spectra for 
    different numbers of satellites  and different observation time. 
The yellow, blue, and black lines correspond to the spectra for 3, 5, and 7 satellites, respectively, and the dashed, dot-dashed, and solid lines correspond to observation time of 1, 100, and 1000 orbits, respectively. 
The {\it lower panel} shows the spectra for different arrangements of 8 satellites and observation time. The red, purple, blue, and black lines are for even spacing, random spacing, logarithmic spacing, and the optimized spacing configurations, and are labeled as (a), (b), (c), and (d), respectively in the legend. The dashed, dot-dashed, and solid lines correspond to results for satellites completing 1, 100, and 1000 orbits flight (8248.7 seconds for each orbit), respectively.
} 
    \label{fig:eigen_sate}
\end{figure}

The eigenvalue spectrum of the matrix $\mathbf{B^{\mr{H}}N^{-1}B}$ are plotted for different array configurations and observation times in Fig.~\ref{fig:eigen_sate}.
We plot the eigenvalue spectra for 3, 5, and 7 satellites,  labeled as (3s), (5s), (7s) in the legend respectively, after 1, 100, and 1000 orbits in Fig.~\ref{fig:eigen_sate} (a). It can be seen from the eigenvalue spectra that for longer observation times and more baselines one would obtain more independent modes with larger eigenvalues. 
In Fig.~\ref{fig:eigen_sate} (b), we show the eigenvalue spectra for different array configurations and the different observation times. 
We find that the logarithmic spacing configuration (c) and the optimized spacing configuration (d) have more modes with high eigenvalues as compared with the even spacing (a) and random spacing (b) configurations. In addition, 
after a period of observation, the optimized spacing configuration results in the highest eigenvalue spectrum among the four configurations.
This suggests that sampling the $uvw$ space for the whole range of scales uniformly is desirable. 

The results of PSF and the eigenvalue spectrum are consistent and complementary to each other,
both quantifying the imaging ability of the array with different designs and observation times.
From the above analysis of the PSF and the eigenvalue spectra, we conclude that the imaging ability of a lunar orbit array can be improved by increasing the number of satellites, extending the observation time, and optimizing the satellite arrangement.
The proposed optimized spacing configuration is the preferred option for the DSL mission, as it results in the sharpest PSF and the lowest noise covariance.

\subsection{Point source sensitivity}
\label{sec:point source sensitivity}
In the above, we have assessed the global map-making capability of a DSL-like array by simulating the full-sky
imaging with $1^\circ$ resolution (HEALPix $N_{\mr{side}} = 64)$.
We now study the sensitivity of the array for point sources by focusing on a small part of sky with higher resolution.

Theoretically, the surface brightness sensitivity of an interferometer array is 
\begin{equation}
T_{\rm RMS} = \frac{L_{\rm max}^2 T_{\rm sys}}{A_{\rm eff} \sqrt{N(N-1)\, t_{\rm obs} \Delta \nu}},  
\label{eq:Trms}
\end{equation}
where $L_{\rm max}$ is the maximum baseline, $A_{\rm eff}$ is the effective area of the antenna, $N$ is the number of interferometric elements, and $t_{\rm obs}$ is the total observation time. The practical sensitivity of DSL may deviate from the theoretical value due to the practical issues involved.
Below, we examine the point source detection from the reconstructed sky map.

We select a dark sky region away from the Galactic plane, for example the region A in Fig. \ref{fig:psf_position}, 
and then apply our synthesis method to obtain the reconstructed sky map with the full resolution of the DSL array for a given frequency. 
We calculate the local brightness temperature fluctuation $T_{\rm RMS}$ in the selected area, which may vary from region to region. Setting the detection threshold at $5 \sigma$, the flux density limit for source detection is
\begin{eqnarray}
S_{\rm min}=\frac{2\, k_{\rm B}}{\lambda^{2}}\,(5\,T_{\rm RMS}) \Omega,
\end{eqnarray}
where $\Omega \approx 1.13\, \theta_{\rm o}^2$ is the beam solid angle, in which
the prefactor of 1.13 is adopted in analogy to a Gaussian beam. However, we do not assume any shape for the beam here, and the beam size in estimating the point source sensitivity is $\theta_{\mr{o}} \approx 1.6 \times \lambda / L_{\rm max}$, which is determined by the full width of half magnitude of the PSF from our simulation, where $L_{\rm max} = 100 \, {\rm km}$ is the longest baseline for the DSL mission.

\begin{figure}
	\centering
	\includegraphics[width=0.95\columnwidth]{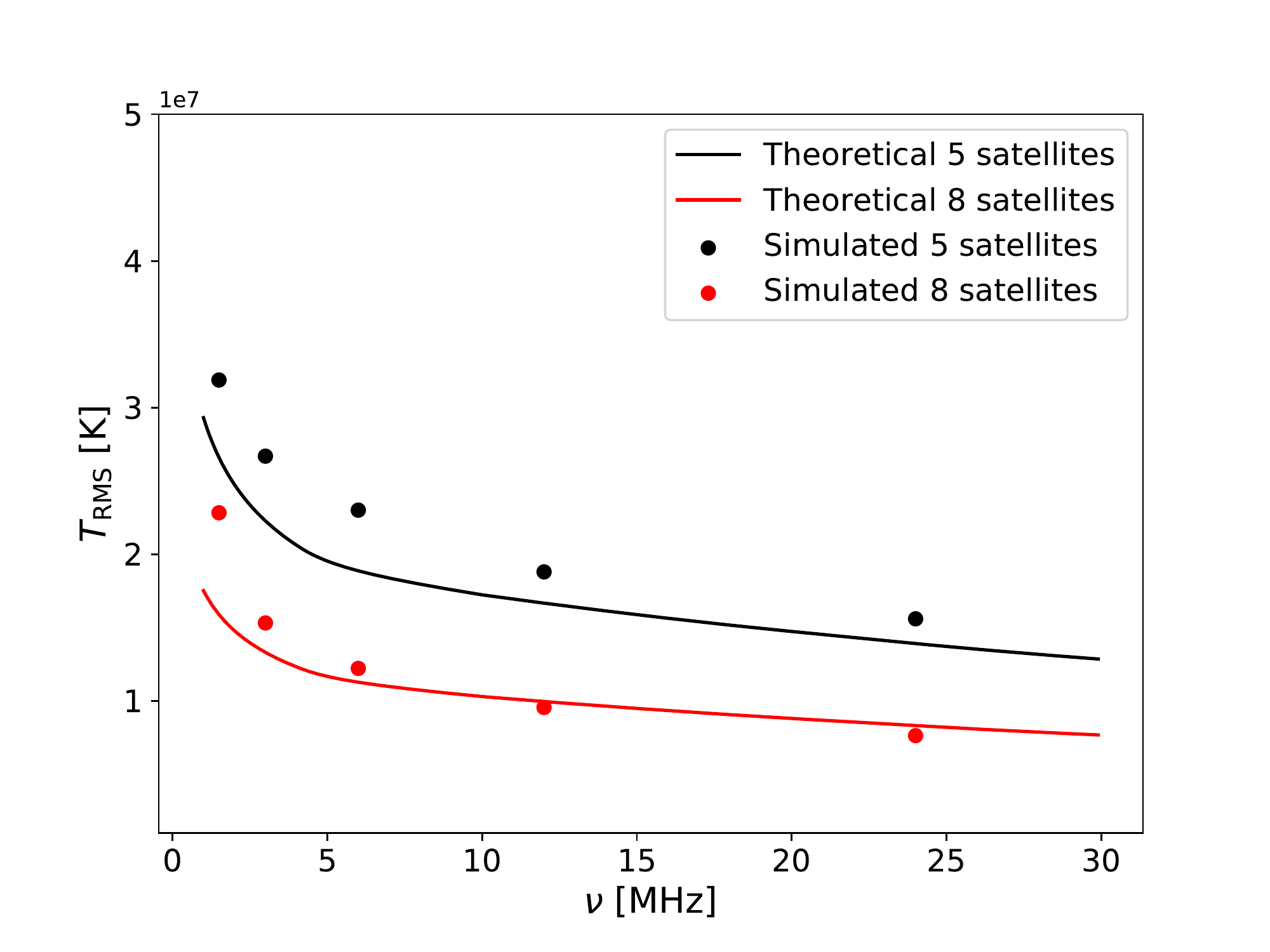}
  \caption{Comparison of the theoretical (black and red lines) and simulated (black and red dots) surface brightness sensitivity of a DSL-like array.}
	\label{fig:theoretical_sensitivity}
\end{figure}

\begin{figure}
	\centering
	\includegraphics[width=0.8\columnwidth]{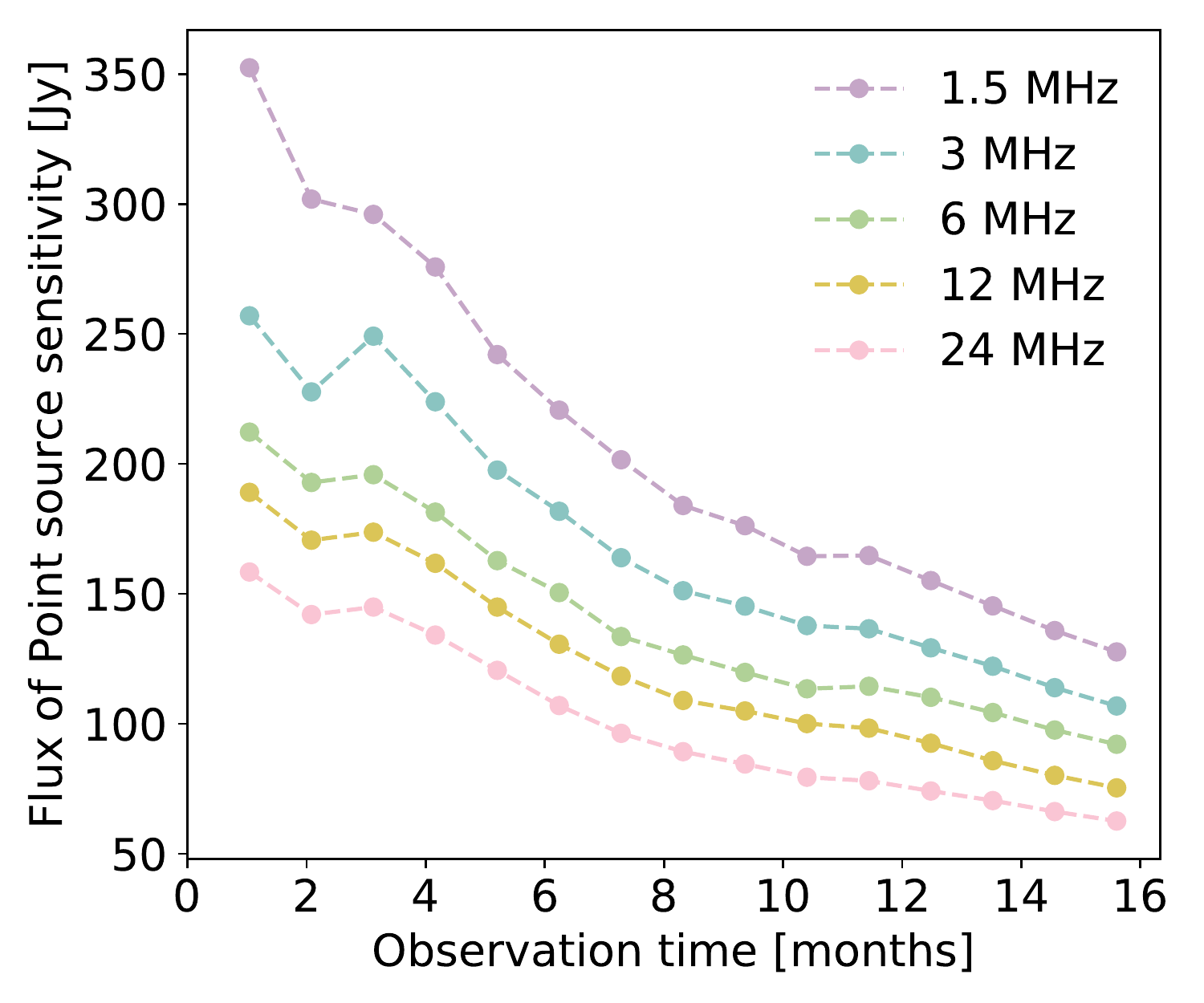}
  	\caption{The 5$\sigma$ flux density limit of a linear array of 5 satellites in optimized spacing configuration as a function of the total observation time at frequencies of 1.5 MHz, 3 MHz, 6 MHz, 12 MHz, and 24 MHz, from top to bottom respectively. The beam sizes of reconstructed sky maps are 11.0, 5.50, 2.75, 1.37, and 0.687 arcmins, respectively. }
	\label{fig:flux_sensitivity}
\end{figure}

We run simulations for a linear array of 5 satellites in optimized spacing configuration at frequencies of 1.5 MHz, 3 MHz, 6 MHz, 12 MHz, and 24 MHz, with the pixel size of 6.87 arcmin, 3.44 arcmin, 1.72 arcmin, 0.859 arcmin, and 0.429 arcmin, respectively, which are a bit finer than the theoretical resolution of the 100 km baseline. We obtained $T_{\rm RMS}$ and the detection threshold of $S_{\rm min}$ according to the simulated beam size $\theta_{\rm o}$, at different frequencies.
As shown in Fig.~\ref{fig:theoretical_sensitivity}, we find that the simulated $T_{\rm RMS}$ generally agrees with the theoretical expectation of Eq.(\ref{eq:Trms}).
The results for the flux density limit are plotted in Fig.~\ref{fig:flux_sensitivity}, showing the increase of sensitivity with the increasing observation time. 
The observing band width is assumed to be 8 kHz according to the current DSL mission concept, which is fairly narrow, but note that the sensitivity may be improved if a larger bandwidth could be achieved with the inter-satellite communication link. The point source sensitivities for different frequencies and different observation times are listed in Table~\ref{tab:part-sky results}.

\subsection{Detectable Sources}

The expected number of detectable sources depends on the luminosity function and the spectral energy distribution of radio sources at ultra-long wavelengths, both of which are completely unknown below 30 MHz. Tentative radio source counts have been obtained at 41.7 MHz and 61 MHz by the LOFAR array, but it is still not sure if the derived shallow spectral index is due to residual calibration issues or the intrinsic property of sources at these frequencies \citep{LOFARsource}. 
Therefore, here we compare three models extrapolated from observations at higher frequencies. 

{\bf 1. VLSS model.} 
\cite{2009NewAR..53....1J} extrapolated the faint-end source number count derived from the VLA Low-Frequency Sky Survey (VLSS) at 74 MHz towards lower frequencies, assuming a spectral index of $\alpha = 0.7$ that is typical for optically thin synchrotron sources. The number density of sources is given by
\begin{equation}
N_{>}(S)=1800\, \mr{deg}^{-2}\left(\frac{S}{10\, \mr{mJy}}\right)^{\beta}\left(\frac{v}{10\, \mr{MHz}}\right)^{-\alpha},
\end{equation}
where S is the flux density limit, measured in mJy, and $\beta = -1.3$. In this model, the source number distribution function $n(S)$ follows a power-law distribution above a certain threshold.

{\bf 2. SSM model.} The all-sky map predicted by the SSM model is used as the input sky map in our simulation. By combining the data from the NRAO VLA Sky
Survey (NVSS) at 1.4 GHz \citep{1998AJ....115.1693C} and the Sydney University Molonglo Sky Survey (SUMSS) at 843 MHz \citep{1999AJ....117.1578B, 2003MNRAS.342.1117M}, the SSM model also includes a point source catalogue covering the entire sky, but given at higher frequencies.
The source distribution function at 1.4 GHz is given by \citep{2019SCPMA..6289511H} 
\begin{eqnarray}
n(S) \approx 1300\, S^{-1.77}.
\end{eqnarray}
The total number of detectable sources in the whole sky is then 
\begin{equation}
    N_{>}(S)=4\pi \int_{S} n(S)\, \mr{d} S.
    \label{eq:source_count}
\end{equation}
Here we extrapolate the source count at 1.4 GHz down to below 30 MHz using a spectral index of $\alpha=0.8157$, which is fitted from sources
in the overlap region of SUMSS and NVSS \citep{2019SCPMA..6289511H}.

{\bf 3. GLEAM model.} We may also use a more up-to-date source count derived from the GaLactic and Extragalactic All-sky MWA survey (GLEAM) at 200 MHz, 154 MHz, 118 MHz and 88 MHz, covering 24,831 $\mr{deg}^2$ \citep{Franzen2019PublAstronSocAust}. We apply a weighted least squares fit to the GLEAM source counts at 154 MHz, with a $5^{\mr{th}}$ order polynomial given by 
\begin{equation}
    \begin{aligned}
        \log \left(S^{2.5} n(S)\right) &= a_{0} (\log S)^{0}+a_{1} (\log S)^{1}+a_{2} (\log S)^{2}\\
        &+a_{3} (\log S)^{3}+a_{4} (\log S)^{4}.
    \end{aligned}
\end{equation}
The fitted parameters are
$a_0 = 3.5230 $, 
$a_1 = 0.2979 $, 
$a_2 = -0.3986 $, 
$a_3 = -0.0204 $, and
$a_4 = 0.0376 $ at 154 MHz \citep{Franzen2019PublAstronSocAust}.
The total number of detectable sources also follows Eq.(\ref{eq:source_count}).
Then the source counts are extrapolated to  1$\sim$30 MHz with a spectral index of $0.8$.

\begin{table*}
	\centering
	\caption{The point source sensitivity of a DSL-like linear array of 5 satellites derived from part-sky imaging simulations. The nine columns are the observing frequency, integration time, resolution of the reconstructed map, the root mean square of sky brightness $T_{\rm RMS}$ and the $5\sigma$ flux density limit after 1 month's observation, the $5\sigma$ flux density limit after 1 month with $1/3$ (shielded from the Earth) and $1/10$ (shielded from both the Earth and the Sun) effective observation time respectively, and $T_{\rm RMS}$ and the $5\sigma$ flux density limit after a whole precession period.}
	\label{tab:part-sky results}
	\begin{threeparttable}
	\begin{tabular}{ccccccccc}
	\toprule
    \multirow{2}{*}{\makecell[c]{$\nu$ \\ {[MHz]}}}& 
    \multirow{2}{*}{\makecell[c]{$t_{\rm int}$ \\ {[s]}}} &
    \multirow{2}{*}{\makecell[c]{$\theta_{\mr{o}}$ \\ {[arcmin]}}} &
    \multicolumn{4}{c}{$t_{\rm obs} =$ 1 month} &
    \multicolumn{2}{c}{$t_{\rm obs} =$ 1.3 years}\cr
    \cmidrule(lr){4-7} \cmidrule(lr){8-9}
    & & 
    &\makecell[c]{$T_{\rm RMS}$\\ {[$10^7$ K]}} &\makecell[c]{$ S_{\rm min} (5\sigma)$ \\ {[$10^2$ Jy]}}
    &\makecell[c]{$S_{\rm min}(5\sigma)$ [$10^2$ Jy]\\($1/3$ $t_{\rm obs}$)}
    & \makecell[c]{$S_{\rm min}(5\sigma)$  [$10^2$ Jy] \\($1/10$ $t_{\rm obs}$)} 
    &\makecell[c]{$T_{\rm RMS}$\\ {[$10^7$ K]}}& \makecell[c]{$S_{\rm min}(5\sigma)$\\ {[$10^2$ Jy]}}\\
    \midrule
    1.5 & 26.26 &	11.0 &	8.8 & 3.5 & 6.1 & 11 & 3.2 & 1.3	\\
    3.0 &	13.13 &	5.50 &	6.4 & 2.6 &	4.5 & 8.1  & 2.7 &	1.1	\\	
    6.0 &	6.56 &	2.75 &	5.3 &	1.9 & 3.3 &  6.1 & 2.3 & 0.92 \\
    12.0 &	3.28 &	1.37 &	4.7 &	1.9 & 3.3 & 6.0 & 1.9 & 0.75	\\
    24.0 &	1.64 &	0.69  &	4.0 &	1.6 & 2.7 & 5.0 & 1.6 &	0.63\\
  
	\hline
	\end{tabular}	
	\end{threeparttable}
\end{table*}

\begin{figure*}
    \subfigure[]{
    \includegraphics[width=0.65\columnwidth]{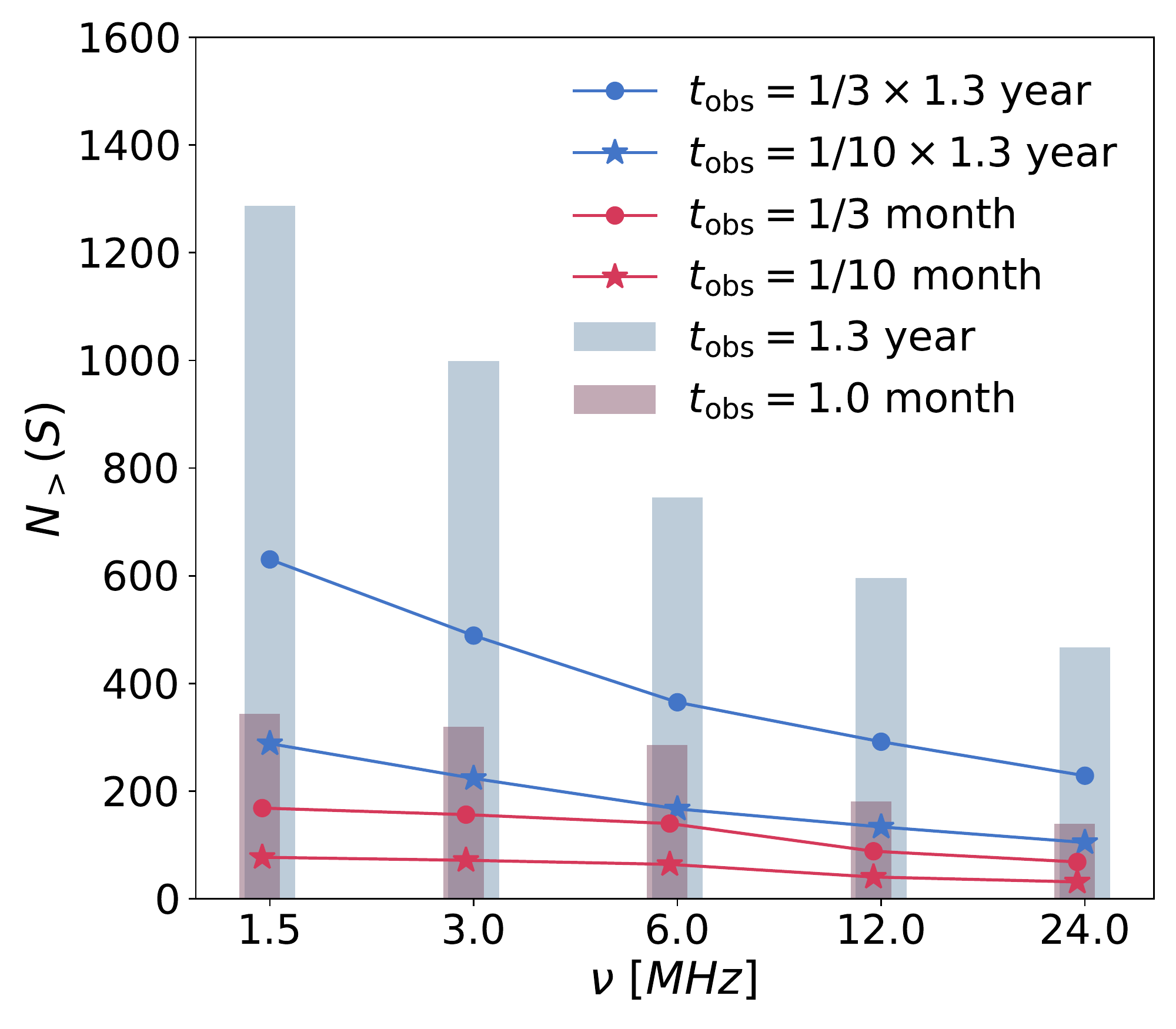}}
    \quad  
    \subfigure[]{
 \includegraphics[width=0.65\columnwidth]{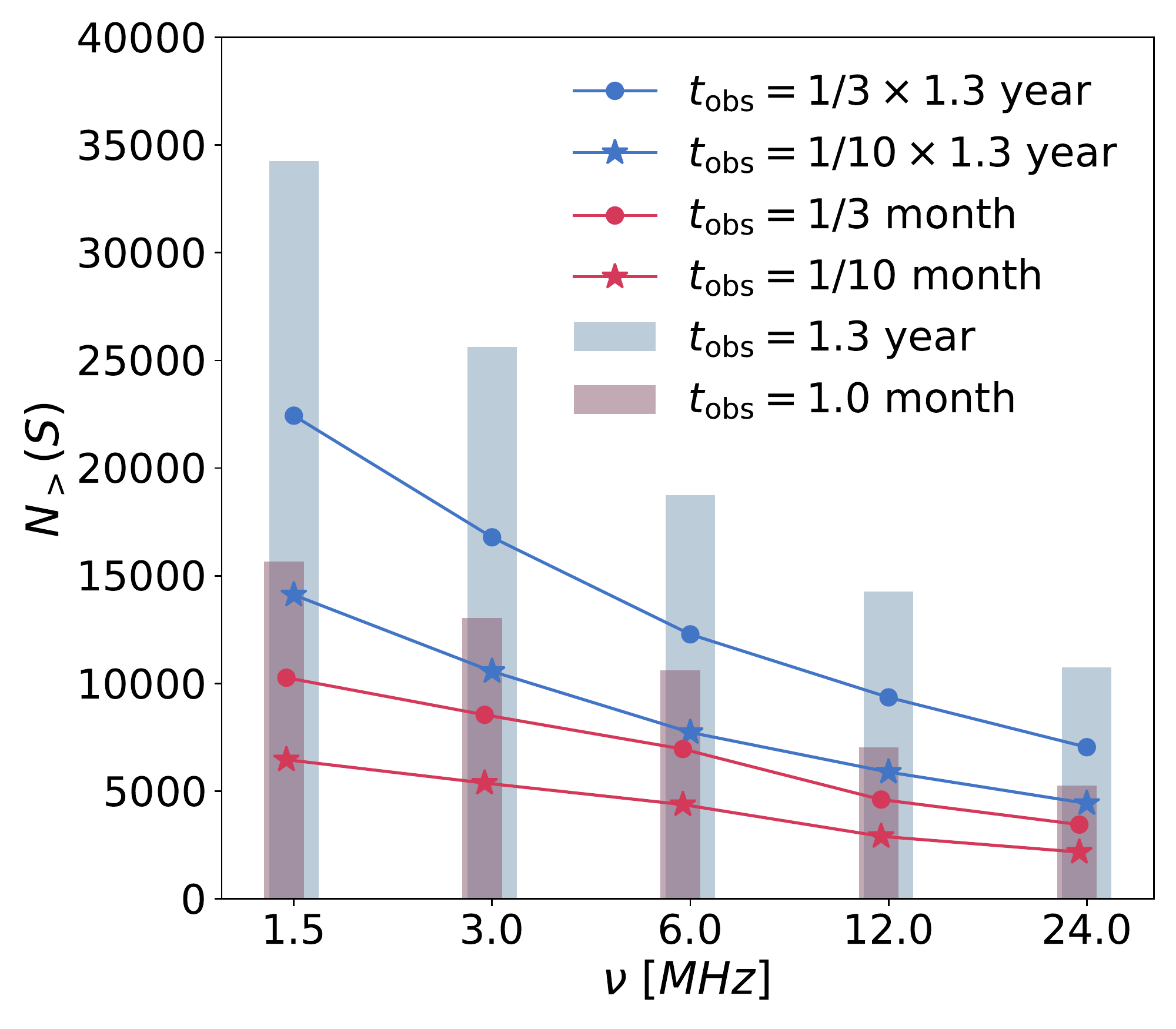}}
    \quad 
        \subfigure[]{
    \includegraphics[width=0.65\columnwidth]{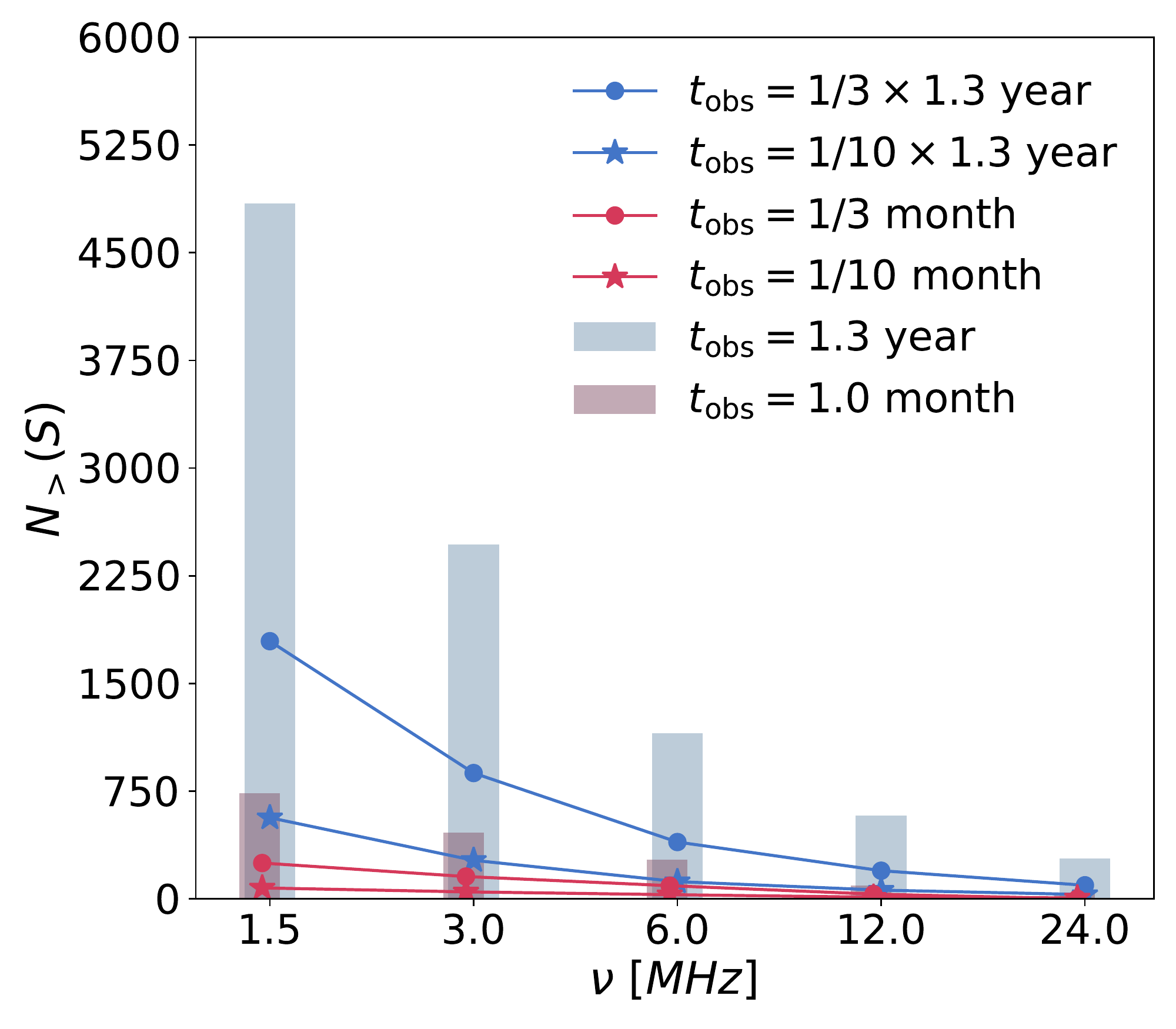}}
    \quad  
    \caption{The number of $5\sigma$ detectable sources in the whole sky  at different frequencies for 5 satellites in the ``optimized spacing" configuration. 
   Left: VLSS model, Middle: SSM model, Right: GLEAM model.  
	}
	\label{fig:sources_frequency}
\end{figure*}

\begin{figure*}
    \subfigure[]{
     \includegraphics[width=0.65\columnwidth]{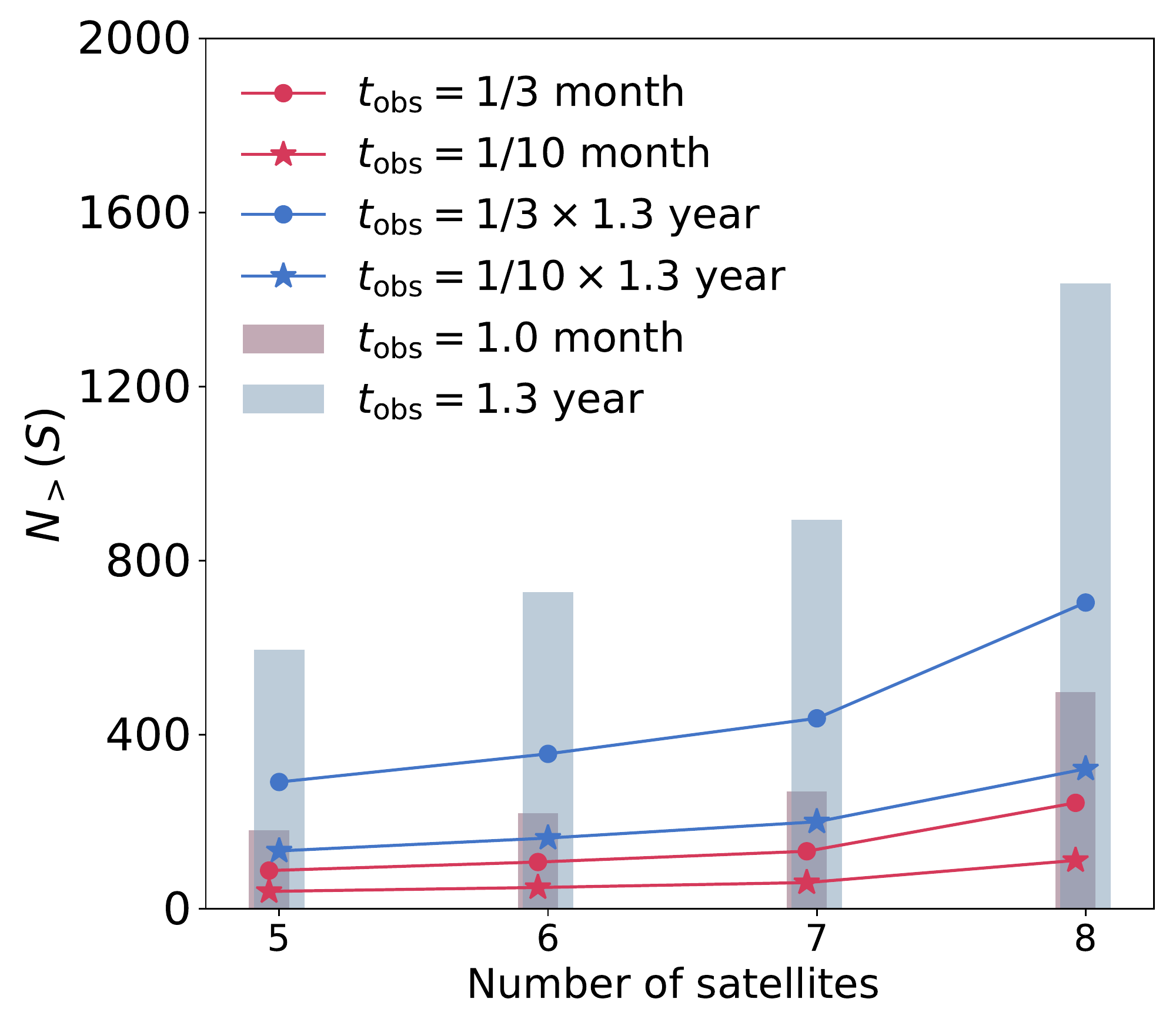}}   
    \quad 
    \subfigure[]{
    \includegraphics[width=0.65\columnwidth]{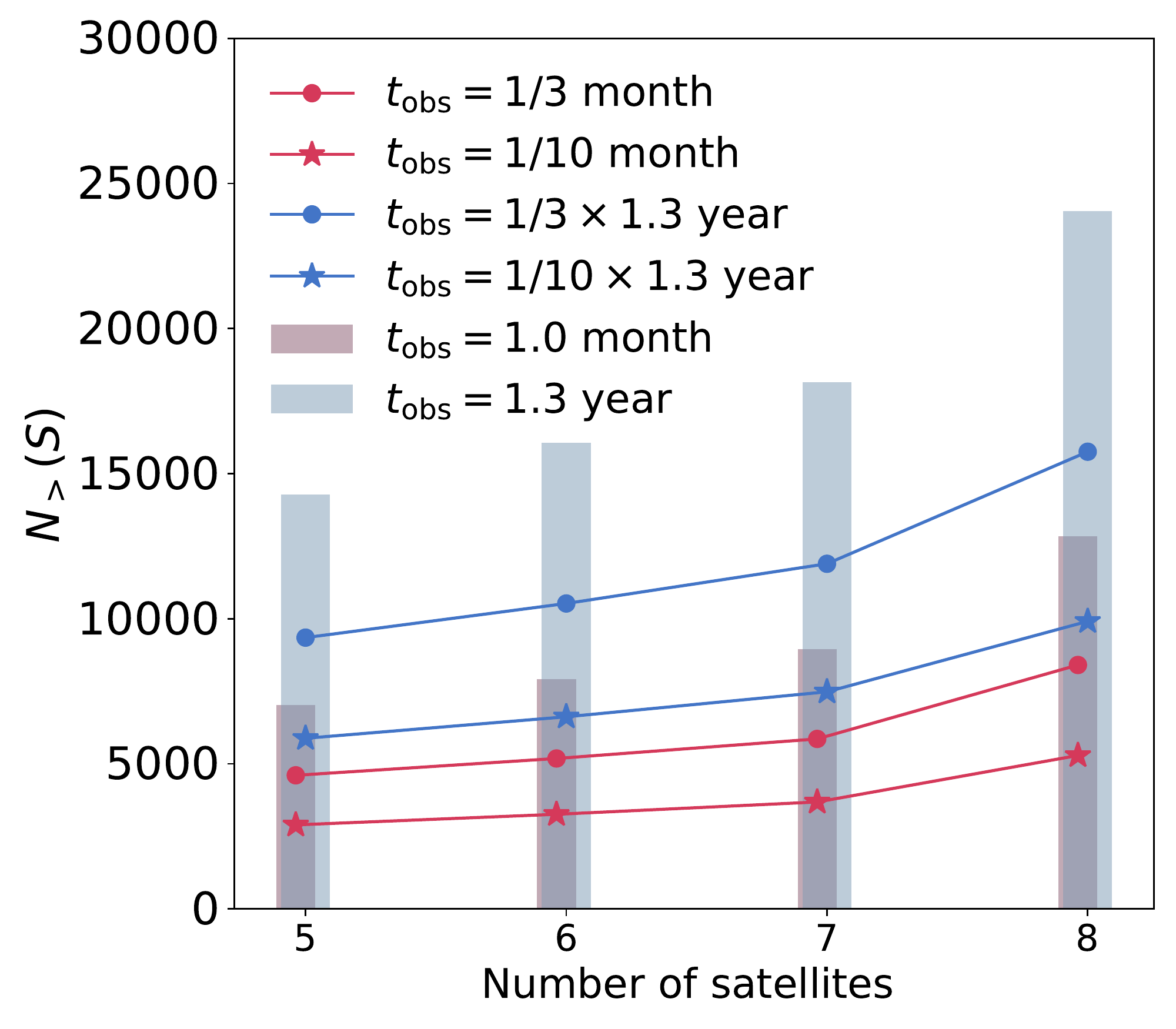}}   
    \quad 
        \subfigure[]{
    \includegraphics[width=0.65\columnwidth]{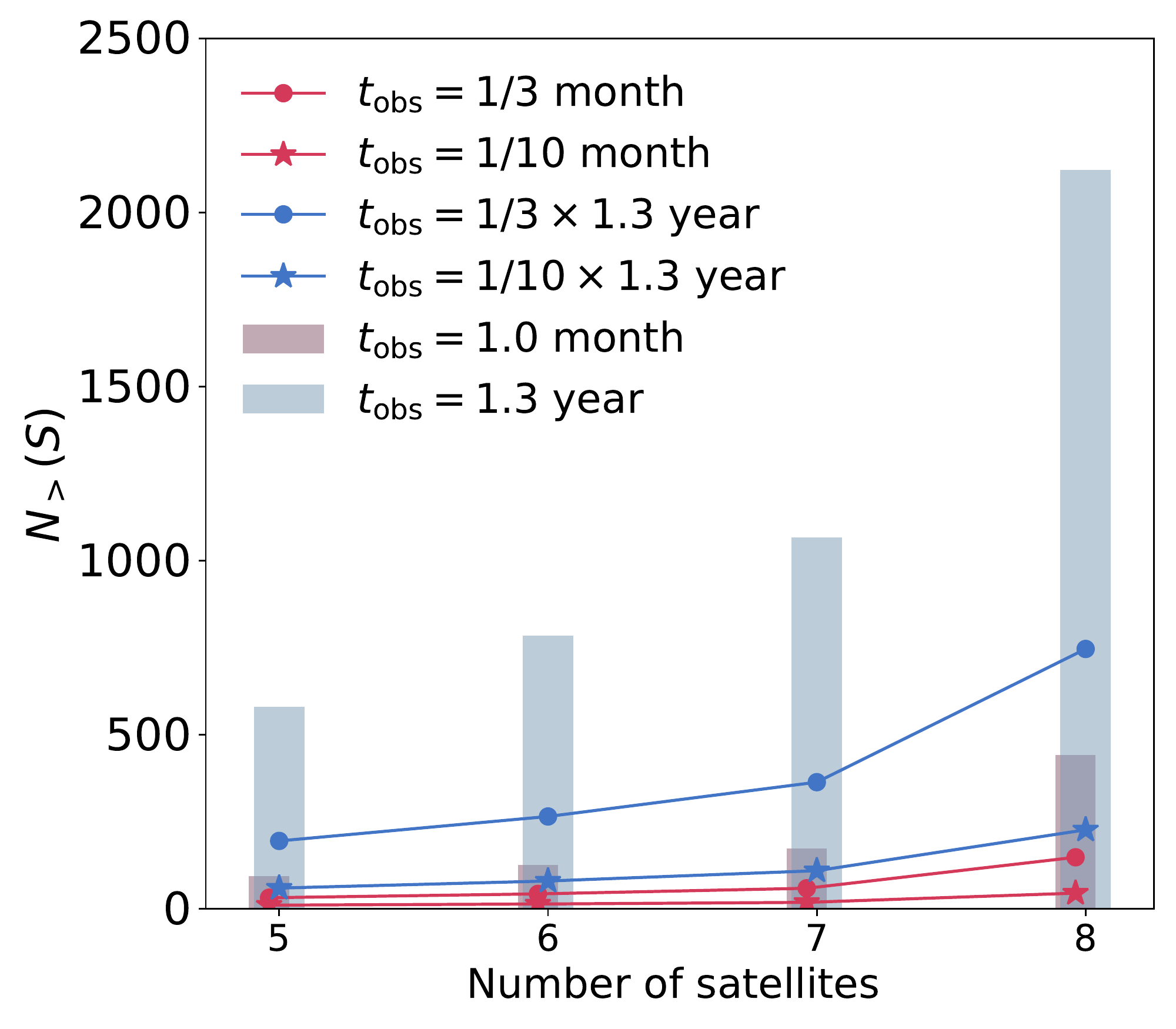}}   
    \quad 
    \caption{ The number of $5\sigma$ detectable sources at 12 MHz for different numbers of satellites in the ``optimized spacing" configuration.  Left: VLSS model, Middle: SSM model, Right: GLEAM model. The red histogram shows the number of point sources detectable in one month's observation, and the blue histogram shows the results for one precession period. The dots and stars with corresponding colors account for the $1/3$ and $1/10$ effective observation time, during which the satellites are shielded by the Moon from the Earth, and from both the Earth and the Sun, respectively. 
	}
	\label{fig:sources_satellite}
\end{figure*}

In Fig.~\ref{fig:sources_frequency} we plot the number of detectable point sources at several frequencies for a linear array of 5 satellites in the ``optimized spacing configuration". The histograms show results for 1 month and 1.3 years of total observation time, without considering the radio frequency interferences from the Sun and the Earth. The predictions for the VLSS model, SSM model and GLEAM model are plotted in the left, middle and right
panels respectively. The number is quite dependent on the source count model, which is currently very uncertain. 
For example, according to the VLSS model, the number of $5\sigma$ detectable sources are 180 (or 590) for one month's observation (or one precession cycle) at 12 MHz, and drops to 140 (or 466) for one month's observation (or one precession cycle) at 24 MHz. 
The GLEAM model predicts more detectable sources at low frequencies, but fewer detectable sources at the high-frequencies, as compared with the VLSS model. According to the GLEAM model, the number of detectable source are 180 (595) sources for one month (one precession cycle) at 12 MHz, but only 44 (280) sources for one month (one precession cycle) at 24 MHz. The SSM model predicts much higher numbers: 10610 (18740) for one month (one precession cycle) at 12 MHz, and 5260 (10700) for one month (one precession cycle) at 24 MHz.

If we only use the data obtained when the Earth is shielded, 
or when both the Earth and Sun are shielded, the available integration time is 
further reduced to1/3 and 1/10 of the total orbital time. The total number of detectable sources would be further reduced. The corresponding results for the VLSS, SSM, and GLEAM models are plotted with dots and stars in the figure. 
In Fig.~\ref{fig:sources_satellite},  we show the number of $5\sigma$ detectable point sources for different numbers of satellites (5 -- 8). The number of detectable sources increases with the number of satellites  as expected.

Note that in the above estimates, we have made several simplifications in our simulations. First, the sky model ignores the absorption effect, which would be important below $\sim 3$ MHz, leading to higher sky temperature and higher flux density limit, underestimating the detectable number of point sources. Secondly, all the above three source models ignore the possible low-frequency turnover in the source spectrum due to synchrotron self-absorption, and the number of sources could therefore be overestimated.  Furthermore, here we have not considered the measurement errors on baselines, the time synchronization errors, the calibration errors on both the amplitude and phase of visibilities, etc. More practical considerations will further decrease the sensitivity and reduce the expected number of detectable sources. 
However, here we have noted the significant variation between the source models, due to the completely unknown nature of sources at the ultra-long wavelengths. A more realistic input sky model and more sophisticated simulations will provide more reliable results of the sensitivity of the DSL array, but the expected number of sources is still quite uncertain.

Above we have discussed the sensitivity limit of point source observation, but whether a source can be detected as a distinct one also depends on the number density of the sources and the angular resolution of the telescope. If multiple sources above the sensitivity limit are located within the same beam, they would appear to be overlapped and indistinguishable, this imposes a confusion limit for point source detection, which usually limit the mean surface density of the point sources to be less than $1/30$ per beam  \citep{2001MNRAS.325.1241V, 2012ApJ...758...23C}.
In the most extreme case of the SSM model, if the observation is performed with 8 satellites at 1.5 MHz, there are less than 30,000 detectable 
sources after one precession cycle using only the Earth-shielded observing time, while there are 
$3 \times 10^6$ pixels for the whole sky at the observational resolution, it amounts to less than one source in 100 pixels.
Note that the results at 1.5 MHz may be severely affected by the free-free absorption effect on both the sky temperature and the source luminosity. At the more reliable frequencies of 12 MHz and 24 MHz, the SSM model predicts about 15,700 and 12,000 detectable sources respectively, after one precession cycle using 8 satellites and the Earth-shielded observing time, while there are $5 \times 10^7$ and $2 \times 10^8$ pixels in the whole sky for 12 MHz and 24 MHz, respectively.
Note this is the sensitivity limit for a single channel with 8 kHz bandwidth. The sensitivity could be significantly improved for the continuum sources by using multi-channel data. Combining the 30 channels to achieve a bandwidth of $30\times 8$ kHz, the 
number of $5\sigma$ sources in the SSM model are then about 58,300 and 45,200 at 12 MHz and 24 MHz respectively,
after a complete precession cycle using 8 satellites and the Earth-shielded observing time. 
Thus the confusion limit is not a serious concern in our case.

\section{Conclusions and Discussions}
\label{sec:discuss}

In this paper, we present the end-to-end simulation for a linear interferometer array on lunar orbit operating at the ultra-long wavelength. We adopt the specific parameters of the DSL mission concept, and for the first time, assessed the imaging capability and obtained the point source sensitivity of a DSL-like array, taking into account the system noise (\S \ref{subsec:noise}), and the various practical issues such as the antenna response (\S \ref{sec:response}) and the time-varying Moon blockage (\S \ref{sec:shade}).

We first generate an input sky map with the SSM model \citep{2019SCPMA..6289511H}, make mock observations by computing the visibilities, taking into account the realistic orbital motion of the satellites, and then apply the imaging algorithm developed by \citet{2018AJ....156...43H} to reconstruct the sky map. By comparing the relative errors, MSE, RS values, and SSIM, we quantify the global imaging capability of the interferometer array of different configurations, and investigate the effects of the various practical issues (Fig.~\ref{fig:full_sky_results2} and Table~\ref{tab:full_sky_results}). We find that, although these various observational and systematic effects have some impacts on the imaging quality, we can still reconstruct the sky map reasonably well with the limited number of satellites. The imaging quality can be improved by increasing the number of satellites, and prolonging the observation time.

In order to optimize the observing strategy, we have made a quantitative comparison of the PSF and the eigenvalue spectrum of the inverse covariance matrix for $3 \sim 8$ satellites, different observation times, and four array configurations. We found that in addition to increasing the number of satellites and observation time, the imaging capability of the array can be further improved by optimizing the array configuration. We have proposed the optimized spacing configuration for a linear array, which can yield both a sharp PSF and high eigenvalue spectrum.

Though we have made some optimizations, the computation for making a full-sky map at high resolution is still prohibitive. After verifying the consistency between the full-sky and part-sky imaging results, we made end-to-end simulation for small regions in the sky with the required resolutions for different frequencies. We derive the flux density threshold for point source detection, and estimate the number of detectable point sources in the whole sky. As we do not yet  know the distribution of sources in this band, the number depends on the model we assumed. According to the more conservative VLSS model, after observing for a whole precession cycle of 1.3 years, we expect more than 590  sources to be detected at 12 MHz, and more than 460 sources at 24 MHz, for a linear array of 5 satellites. The numbers increases to more than 1430 and 750 respectively if we could have an array of 8 satellites. However, if we consider only $\sim 1/10$ effective observation time when the satellites are shielded from both the Earth and the Sun, then the detectable source number reduces to about 130 (12 MHz) and 104 (24 MHz) for 5 satellites, and about 320 (12 MHz) and 160 (24 MHz) for 8 satellites. These numbers are however simple extrapolations from a few observations of much higher
frequencies and are subject to large uncertainties. In general, we expect the number of sources detectable by the DSL fall in the $10^2 \sim 10^4$ range.

The present study provide a first evaluation of its imaging sensitivity and quality of the upcoming lunar orbit array by end-to-end simulation. We show that  high quality images can be obtained despite some practical issues which may affect the observation.
There are more systematic effects remain to be investigated, such as the measurement errors on the baselines and time synchronization, the calibration errors, the reflection by the Moon 
surface, and the residual RFIs from the satellites themselves, etc. We plan to incorporate these into the simulation in future works.

\section*{Acknowledgements}
We thank Bin Yue for helpful discussions.
This work is supported by 
the Chinese Academy of Sciences (CAS) Strategic Priority Research Program XDA15020200,
the National Natural Science Foundation of China (NSFC) grant 11973047, 11633004, the National Key R\&D Program of China (2020YFE0202100), 
the National SKA Program of China No. 2020SKA0110401,
the Ministry of Science and Technology (MoST) inter-government cooperation program China-South Africa Cooperation Flagship project 2018YFE0120800, the CAS Frontier Science Key Project QYZDJ-SSW-SLH017, and the China Manned Space Project with NO. CMS-CSST-2021-B01.
This work used resources of China SKA Regional Centre prototype \citep{2019NatAs...3.1030A} funded by the National Key R\&D Programme of China (2018YFA0404603) and Chinese Academy of Sciences (114231KYSB20170003).

\section*{Data Availability}
The data underlying this article will be shared on reasonable request to the corresponding author.

\bibliographystyle{mnras}

\bibliography{references}

\bsp
\label{lastpage}
\end{document}